%% file: ms.tex
\documentclass[useAMS, usenatbib]{mnras}  
\usepackage{color}
\usepackage{booktabs}   
\usepackage{tabularx}
\usepackage{float}
\usepackage{totcount}
\usepackage{cuted}

\usepackage[dvipsnames, table]{xcolor}
\usepackage{graphicx}
\usepackage{multirow}
\usepackage[section]{placeins}
\usepackage{subfig}
\usepackage{multicol}
\usepackage{widetext}
\usepackage{upgreek}    
\usepackage{enumitem}   
\setlist[enumerate]{leftmargin=*}   
\setlist[itemize]{leftmargin=*}   

\usepackage{lipsum} 
\usepackage{amsmath}
\usepackage{amssymb}

\usepackage{txfonts}     

\graphicspath{ {figs/} }

\input{AstroComs}                   
\input{JAbs}                        

\newcommand{\frest}{f_\trt{r}}
\newcommand{\fobs}{f}

\newcommand{\hc}{h_\trt{c}}
\newcommand{\hs}{h_\trt{s}}

\newcommand{\hcs}{h_\trt{c,s}}

\newcommand{\volfactor}{\Lambda_{ij}}
\newcommand{\poisson}{\mathcal{P}}

\newcommand{\frefill}{\mathcal{F}_\trt{refill}}

\newcommand{\mchirp}{\mathcal{M}}     
\newcommand{\volill}{V_\trt{ill}}   

\newcommand{\bigt}{\scalebox{1.2}{\ensuremath{\uptau}}}

\newcommand{\tgw}{\bigt_\tr{GW}}

\newcommand{\fgw}{f_\tr{GW}}

\newcommand{\eccinit}{e_0}

\newcommand{\forefac}{\lambda_\textrm{fore}}
\newcommand{\hcfore}{h_c^\textrm{fore}}
\newcommand{\hcback}{h_c^\textrm{back}}
\newcommand{\rnamp}{A_\trt{RN}}

\newcommand{\freqper}[1]{1/\left(#1 \, \yr\right)^{-1}}
\newcommand{\rnind}{{\protect\scalebox{1.1}{\ensuremath{\gamma{}}}_\trt{RN}}}

\newcommand{\wnrms}{\sigma_\trt{WN}}
\newcommand{\micros}{\mu \textrm{s}}
\newcommand{\nanos}{\textrm{ns}}

\defcitealias{hkm09}{HKM09}
\defcitealias{bbr80}{BBR80}
\defcitealias{paper1}{Paper-I}
\defcitealias{paper2}{Paper-II}
\defcitealias{rsg15}{RSG-15}

\newtotcounter{citnum} 
\def\oldbibitem{} \let\oldbibitem=\bibitem
\def\bibitem{\stepcounter{citnum}\oldbibitem}

\title[PTA GW Singles]{Single Sources in the Low-Frequency Gravitational Wave Sky:\\properties and time to detection by pulsar timing arrays}
\author[L.Z.~Kelley et al.]{
    \hspace{-0.054in}Luke Zoltan Kelley$^{1}$\thanks{E-mail:lkelley@cfa.harvard.edu},
	Laura Blecha$^{2}$,
    \newauthor
	Lars Hernquist$^{1}$,
	Alberto Sesana$^{3}$,
	Stephen R. Taylor$^{4}$
    \vspace{0.1in} \\
    $^{1}$ Harvard University, Center for Astrophysics \\
    $^{2}$ University of Maryland \\
    $^{3}$ University of Birmingham \\
    $^{4}$ Jet Propulsion Laboratory, California Institute of Technology
}

\begin{document}

\pagerange{\pageref{firstpage}--\pageref{lastpage}} \pubyear{2017}

\maketitle
\label{firstpage}

\begin{abstract}
    We calculate the properties, occurrence rates and detection prospects of individually resolvable `single sources' in the low frequency gravitational wave (GW) spectrum.  Our simulations use the population of galaxies and massive black hole binaries from the Illustris cosmological hydrodynamic simulations, coupled to comprehensive semi-analytic models of the binary merger process.  Using mock pulsar timing arrays (PTA) with, for the first time, varying red-noise models, we calculate plausible detection prospects for GW single sources and the stochastic GW background (GWB).  Contrary to previous results, we find that single sources are at least as detectable as the GW background.  Using mock PTA, we find that these `foreground' sources (also `deterministic'/`continuous') are likely to be detected with $\sim 20 \, \yr$ total observing baselines.  Detection prospects, and indeed the overall properties of single sources, are only moderately sensitive to binary evolution parameters---namely eccentricity \& environmental coupling, which can lead to differences of $\sim 5 \, \yr$ in times to detection.  Red noise has a stronger effect, roughly doubling the time to detection of the foreground between a white-noise only model ($\sim 10$ -- $15 \, \yr$) and severe red noise ($\sim 20$ -- $30 \, \yr$).  The effect of red noise on the GWB is even stronger, suggesting that single source detections may be more robust.  We find that typical signal-to-noise ratios for the foreground peak near $f = 0.1 \, \pyr$, and are much less sensitive to the continued addition of new pulsars to PTA.
\end{abstract}

\begin{keywords}
quasars: supermassive black holes, galaxies: kinematics and dynamics
\end{keywords}

\section{Introduction}

    Pulsar timing arrays (PTA) can measure gravitational waves (GW) by precisely measuring the advance and delay in pulses from galactic millisecond pulsars \citep{hellings1983}.  The absence of deviations in the timing of a single pulsar can be used to place constraints on the presence of GW signals \citep{estabrook1975, sazhin1978, detweiler1979, bertotti1983, blandford1984} while the cross-correlation of timing data from an array of pulsars can be used to directly measure metric deviations \citep{foster1990, jenet2005, yardley201102, demorest2013}.  The expected sources pf detectable GW in the low-frequency PTA regime ($f \sim 1 \textrm{ yr}^{-1} \sim 10 \textrm{ nHz}$) are massive black hole binaries (MBHB; $M_\mathrm{tot} \sim 10^6 - 10^{10} \, \msol$) in stable orbits, typically millions of years before coalescence \citep{rajagopal1995, wyithe2003, jaffe2003, sesana2004, enoki2004}.

    Three independent PTA are in operation: the European \citep[EPTA,][]{kramer2013, desvignes2016}, NANOGrav \citep{mclaughlin2013, arzoumanian2015a}, and Parkes \citep[PPTA,][]{manchester2013a, reardon2016}.  The International PTA \citep[IPTA,][]{hobbs2010,Verbiest201602} is a collaboration between all three which uses their combined data to boost sensitivity.  Prospects for GW detection by PTA depend sensitively on the continued discovery of additional, low-noise, millisecond pulsars to incorporate into the networks \citep[e.g.][]{taylor2016}.  Additionally, as observing baselines on existing pulsars increase, the presence of red-noise in pulsar timing residuals can have ever increasing and more dominant effects.  The source of red-noise, or even the relative contribution of astrophysical versus instrumental origins, is unclear \citep[see, e.g.][]{caballero2016}.

    \begin{figure}
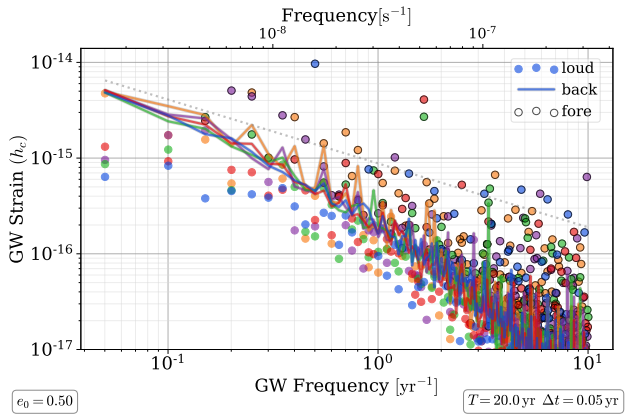

    \centering
    \includegraphics[width=\columnwidth]{{{fig1}}}
    \caption{Five realizations of the low-frequency GW sky are shown from our models based on black holes from the Illustris simulations.  The GW spectrum is separated into the loudest individual sources per frequency-bin (circles) and the remaining `background' of all other systems (lines).  Single sources which are louder than the background are highlighted (black circles), and constitute the GW `foreground'.  The power-law GWB spectrum (\refeq{eq:gwb_plaw}), assuming a continuum of sources evolving purely due to GW emission is shown with the dotted grey line.}
    \label{fig:gwb_singles_zoom}
    \end{figure}

    The low-frequency GW sky can be classified in terms of a stochastic GW Background (GWB)---the superposition of many unresolved GW sources; and deterministic GW sources, where individual binaries are resolvable above the background.  We refer to these resolvable single sources as the GW Foreground\footnote{Note that the term `foreground' is also sometimes used in the literature to refer to general astrophysical sources of GW, at times even the GWB.  Resolvable single sources are also referred to as `continuous' and  `deterministic' sources.} (GWF).  Figure \ref{fig:gwb_singles_zoom} shows five randomly selected realizations of GW signals from our models, decomposed into the loudest single-source in each frequency bin (circles) and the background (all other sources combined; lines).  `Foreground' sources with strains larger than that of the background are highlighted in black.  At low frequencies, the GWB characteristic-strain tends to follow a power-law \citep{phinney2001},
    \begin{equation}
    \label{eq:gwb_plaw}
    \hc \propto \ayr \scale[-2/3]{f}{1 \, \pyr},
    \end{equation}
    with an amplitude, typically referenced at a frequency $f = 1\,\pyr$, on the order of $\ayr \sim 10^{-15}$ \citep{wyithe2003}.  The power law model is based on the assumption of a continuum of sources, evolving purely due to GW emission and over an indefinite frequency range.  The $-2/3$ spectral index emerges based on the rate at which the binary separation `$a$' shrinks (`hardens') due to GW radiation \citep[see, e.g., the derivation in][]{sesana2008}.  At small separations the residence time, $a / (da/dt)$, decreases enough that the average number of binaries emitting in a given frequency bin reaches order unity, causing the spectrum to steepen and falloff from the power-law prediction \citep{sesana2008}.

    While neither class of low-frequency GW signal has been observed, the most recent upper-limits on a GWB \citep{lentati2015, Arzoumanian201508, shannon2015, Verbiest201602} are now astrophysically constraining, if not surprising \citep[e.g.][]{paper1, middleton2017}.  Additionally, some groups have even begun to constrain the presence of individual MBHB in nearby galaxies \citep[e.g.][]{babak2015, Schutz201510}.

    Recent studies predict that the IPTA could detect the GWB by about 2030, if fiducial parameters are reasonable \citep[e.g.][]{taylor2016, paper2}.  It is often stated that the GWB is expected to be detected before the foreground.  However, only a few papers exist with quantitative predictions for single-source rates \citep{sesana2009, ravi2014}\footnote{\citet{roebber2016} do not discuss single-sources \textit{per se}, but much of the analysis is relevant, and their Fig.~2 of a high-resolution realization of GW sources, is very informative.}, and only one which includes calculations of detection probabilities \citep{rsg15}.  \citet{sesana2009} use dark matter halos and merger rates from the Millennium simulations combined with a variety of observational BH--galaxy scaling relations to calculate resolvable MBHB strains.  They predict at least one system with residuals between $\sim 5$ -- $50 \, \nanos$ after observations $T = 5\, \yr$ duration, with resolvable sources tending to come from redshift $z \sim 0.3$ -- $1$, and chirp-masses $\log (\mchirp/\msol) \sim 8.5$ -- $9.5$.  \citet{ravi2014} use observationally determined galaxy merger rates and MBH--galaxy scaling relations to construct semi-analytic MBHB systems.  At $\fgw = 1 \, \pyr$ (with $T = 10 \, \yr$), they expect roughly one source with a strain above $10^{-16}$, and a probability of $10^{-2}$ -- $10^{-1}$ for strains above  $10^{-15}$.  Using an IPTA-like model, \citet{rsg15} find a single-source detection probability of $\sim 10\%$ ($\sim 20\%$) after $20 \, \yr$ ($30 \, \yr$) of observations, and overall a $\sim 5$ -- $25\%$ chance that a single-source will be detected before the GWB.

    Single sources offer a largely independent window into MBHB populations and their evolution.  While the amplitude of the GWB, for instance, suffers from degeneracies between the distribution of MBH masses and their coupling to environmental hardening mechanisms, the observation of deterministic sources could break that degeneracy and possibly demonstrate MBHB orbital evolution in real time.  Additionally, unlike the GWB, foreground sources immediately offer the prospect of observing electromagnetic counterparts.  Numerous existing surveys have already identified candidate MBHB systems based on spectroscopic and photometric techniques, for example searching for periodic variability in the CRTS \citep{charisi2016} and PTF surveys \citep{graham2015}.  While there are signs that a large fraction of these candidates could be false positives \citep{sesana201703}, they are still promising for the possibility of multi-messenger observations.

\section{Methods}
    \label{sec:methods}

    \subsection{MBH Binary Population and Evolution}

        In this section we summarize the key aspects of our methodology which are described in detail in \citet{paper1} \& \citet{paper2}.  We use the galaxies and black hole populations obtained from the Illustris simulations, which coevolve hydrodynamic gas cells along with star, dark matter, and black hole particles over cosmic time \citep{vogelsberger2014b, genel2014, torrey2014, nelson2015}.  Once a galaxy grows to a halo mass of $\sim 7\E{10} \, \msol$ ($M_\trt{star} \sim 10^9 \, \msol$) it is given a MBH with a seed-mass of $10^5 \, \msol$, which then accretes matter from the neighboring gas cells as the galaxy evolves \citep{vogelsberger2013, sijacki2015}.  After a galaxy merger, if two MBH particles come within a gravitational smoothing length of one another (typically $\sim$ kpc), they are manually `merged' into a single MBH with their combined masses.

        We identify these \textit{pseudo}-mergers in Illustris and further evolve the constituent MBH in semi-analytic, post-processing simulations to model the effects of the small scale (sub-kiloparsec) `environmental' processes which mediate the true, astrophysical merger process.  To do this, we calculate density and velocity profiles from each MBHB host galaxy, and use them to calculate hardening rates from dynamical friction, stellar scattering, viscous drag (from a circumbinary disk), and gravitational wave emission.  Eccentric binary evolution can also be included, in which the eccentricity is enhanced by stellar scatterings and decreased by gravitational wave emission.  All binaries are initialized to a uniform, fixed value of eccentricity ($e_0$), which, along with the binary separation, is then numerically integrated in time until each system reaches redshift $z = 0$.  For circular evolution models, we use a scatter prescription based on \citet{magorrian1999}, in which we can vary the effectiveness of scattering based on a parameter\footnote{This `refilling' parameter interpolates between a sparsely filled loss-cone (the region of stars in parameter space able to interact with the binary), and a full one.} $\frefill \in [0.0, \, 1.0]$ --- which acts as a proxy for the overall degree of environmental coupling \citep[see][]{paper1}.

        By associating the presence of each binary in the simulation volume ($80 \, \mathrm{Mpc}^3$, at $z = 0$) with the result of a Poisson process, we can calculate GW spectra from an arbitrary number of observational `realizations' by re-drawing from the appropriate distribution and weighting each binary accordingly.  Specifically, we define a \textit{representative volume factor} for binary $i$ at a time-step $j$,
        \begin{equation}
    	\volfactor = \frac{1}{\volill}\frac{dV_c(z_{ij})}{dz_{ij}} \Delta z_{ij},
    	\end{equation}
        where the comoving volume element is,
        \begin{equation}
    	dV_c(z) = 4\pi \left(1+z\right)^2 \frac{c}{H(z)} \, d^2_c(z) \, dz.
    	\end{equation}
        Here, $H(z)$ is the Hubble constant at redshift $z$ (and corresponding comoving distance $d_c$), $c$ is the speed of light, and $\Delta z_{ij}$ is the redshift step-size for this binary and time step.  $\volfactor$ is the expectation value for the number of \textit{astrophysical} binaries in the past light-cone between redshifts $z_{ij}$ and $z_{ij} + \Delta z_{ij}$, corresponding to each \textit{simulated} binary \& time-step.  To construct an alternate realization, we can draw from the Poisson distribution, $\poisson(\volfactor)$, for each binary and step.  For example, the characteristic GW strain spectrum from all binaries is,
        \begin{equation}
        \label{eq:strain_mc_sum}
        \hc^2(f) = \sum_{ij} \poisson(\volfactor) \sum_{n=1}^\infty \left[ \frac{f}{\Delta f} \, \hs^2(\frest) \left(\frac{2}{n}\right)^2 g(n,e) \right]_{\frest = \fobs(1+z)/n}.
        \end{equation}
        Equation~\ref{eq:strain_mc_sum} takes into account that an eccentricity $e$ leads to a redistribution of GW energy to each harmonic $n$ of the rest-frame orbital frequency $f_r$.  The amount of power observed at $f = f_r \, n / (1+z)$ is given by the power distribution function $g(n,e)$ \citep[see, Eqs.~A1 in][]{paper2}.  For a chirp-mass $\mchirp = \left(M_1 M_2\right)^{3/5} / \left(M_1 + M_2\right)^{1/5}$, the GW strain from a binary with zero eccentricity is,
        \begin{equation}
        \label{eq:source_strain}
        \hs(\frest) = \frac{8}{10^{1/2}} \frac{\left(G\mchirp\right)^{5/3}}{c^4 \, \comdist}
            \left(2 \pi \frest\right)^{2/3},
        \end{equation}
        which, for a circular binary, is emitted entirely at the $n=2$ harmonic.  The \textit{characteristic} strain for a particular source, which takes into account the number of cycles viewed in-band, is,
        \begin{equation}
        \label{eq:source_char_strain}
        \hcs^2(\frest) = \hs^2(\frest) \left(\frac{f}{\Delta f}\right).
        \end{equation}
        Alternatively, the observed timing residual can be calculated directly\footnote{$\delta t \propto \hs \tgw \, N^{1/2}$, where the observed GW period $\tgw = 1/f$, and the number of observed cycles $N = f T$.} as \citep{sesana2009},
        \begin{equation}
        \label{eq:residuals}
        \delta t = \frac{10^{1/2}}{15} \hs(f) \, \left(\frac{T}{f}\right)^{1/2}.
        \end{equation}

    \subsection{Detection Statistics and Pulsar Timing Array Models}
        \label{sec:meth_ds}

        In our analysis we only consider the single loudest sources in each frequency bin as candidates for the foreground.  \citet{ravi2012} point out that PTA can resolve spatially in addition to chromatically, allowing multiple loud sources to be simultaneously extracted from the same bin.  \citet{boyle2012} and \citet{babak2012} demonstrate that this is possible if there are roughly six or more pulsars equally dominating PTA sensitivity.  The uncertainty introduced by neglecting this effect is small compared to those of the models being used, and additionally, based on our results, it is rare for multiple single-sources to each produce comparable strains while also resounding over the GWB.

        Methods for the detection of single sources by existing PTA have been rapidly developed in the last decade \citep[e.g.][]{corbin2010, lee201103, boyle2012, babak2012, ellis201204, ellis2013, taylor201406, zhu201502, zhu201606, taylor201505}.  For detection statistics with mock PTA, we use the methods presented in \citet[][hereafter, \citetalias{rsg15}]{rsg15}.  The statistics for the background, based on cross-correlations between pulsars, were also used in \citet{paper2}.  In this study, however, we explore the distinction between a \textit{true} background with the loudest single-sources per bin removed (`back') at the same time as \textit{all} GW sources included (`both').

        The formalism for single-sources (GWF), based on excess power recovery, requires all pulsars to have the same frequency sampling determined by the observing duration $T$ (which we vary) and cadence $\delta t$ (fixed to $\delta t = 0.05 \, \yr$).  We use Eq.~35~\&~36\footnote{Compared to \citetalias{rsg15}, we define the GW phase after a time $T$ to be $\Phi_T = 2\pi f + \Phi_0$, for an initial phase $\Phi_0$.  We find a slightly different expression for the signal-to-noise after integrating their Eq.~36 than they present in Eq.~46, but the differences are negligible once incorporated into a full analysis.} (\citetalias{rsg15}) to calculate the signal-to-noise ratio (SNR) for each pulsar and each frequency bin, given a GW spectrum and PTA configuration.  Using Eq.~32 (Ibid.) the SNR are converted to bin/pulsar detection probabilities (DP), and Eq.~33 (Ibid.) finally converts to an overall DP for the PTA to detect at least one GWF source.  The single-source detection statistics we use are not designed for eccentric systems, where the signal from an individual source can be spread over multiple frequency bins.  When this happens, we are effectively treating the portion of the signal in each bin as independent, which is obviously sub-optimal.  As most of the binaries contributing to the GW signals have fairly low eccentricity \citep[see, e.g.~\figref{fig:gw-props_ecc}; and][]{paper2}, this should only have a minor effect.

        To calculate plausible detection probabilities, we use mock PTA with a variety of noise models.  We characterize the noise with three parameters, a white-noise amplitude $\wnrms$, and a red-noise amplitude \& spectral index $\rnamp$ \& $\rnind$ for a power-spectrum,
        \begin{equation}
        S_\trt{RN} = \frac{\rnamp^2}{12\pi^2} \left(\frac{f}{f_\trt{ref}}\right)^{\rnind} \, f_\trt{ref}^{-3}.
        \label{eq:red-noise}
        \end{equation}
        Defined in this way, $\rnamp$ corresponds to the equivalent strain at the reference frequency $f_\trt{ref}$, which is always set to $1 \, \pyr$ in our analysis.  All pulsars in a given array use the same noise parameters.

        \begin{table}
        \begin{center}
        \begin{tabular}{c c c c c}
        name    & $h_c^N(f = 0.1 \, \yr)$    & $\wnrms$         & $\rnamp$   & $\rnind$ \\ \hline
        $\textrm{\textbf{a}}$     & $\mathbf{1.0\E{-15}}$               & $0.3 \, \micros$ & -          & -        \\
        $\textrm{b}$     & $1.8\E{-15}$               & $0.3 \, \micros$ & $4\E{-15}$ & $-3.0$   \\
        $\textrm{c}$     & $6.7\E{-15}$               & $0.3 \, \micros$ & $1\E{-15}$ & $-4.5$   \\
        $\textrm{\textbf{d}}$     & $\mathbf{1.8\E{-14}}$               & $3.0 \, \micros$ & $4\E{-15}$ & $-3.5$   \\
        $\textrm{\textbf{e}}$     & $\mathbf{4.6\E{-14}}$               & $3.0 \, \micros$ & $2\E{-13}$ & $-1.5$   \\
        $\textrm{f}$     & $2.0\E{-13}$               & $0.3 \, \micros$ & $2\E{-13}$ & $-3.0$   \\
        \end{tabular}
        \caption{Parameters of the noise models used in our mock PTA.  The 2nd column gives the total noise, in units of characteristic strain, at $f = 0.1 \, \pyr{}$.  The remaining columns are the white-noise RMS $\wnrms{}$, and the red-noise amplitude \& spectral index, $\rnamp{}$ \& $\rnind{}$ (see: \refeq{eq:red-noise}).  The parameters for these models were chosen manually while trying to accurately represent the properties of observed pulsars (see: \figref{fig:noise}).  Much of our analysis focuses on models `a', `d' \& `e', which are highlighted.}
        \label{tab:noise}
        \end{center}
        \end{table}

        Noise models were selected to cover a parameter space comparable to that measured by PTA.  The noise characterization between different PTA can vary significantly, however, at times being inconsistent.  For example, the red noise of \texttt{J1713+0747} is characterized by an amplitude and spectral index: $\rnamp = 2\E{-15}$ \& $\rnind = -4.8$ by the EPTA \citep{caballero2016}, and $\rnamp = 3.5\E{-14}$ \& $\rnind = -2.0$ by Parkes \citep{reardon2016}.  Similarly, \texttt{J1910+1256} is given $\rnamp = 2.8\E{-13}$ \& $\rnind = -1.9$ by NANOGrav \citep{arzoumanian2015a}, but $\rnamp = 2.9\E{-15}$ \& $\rnind = -5.9$ in the IPTA data release \citep{Verbiest201602}.  The parameters of the six noise models we use are given in Table~\ref{tab:noise}, and are plotted against pulsars from each PTA in \figref{fig:noise}.  Much of our analysis focuses on noise models `a', `d' \& `e', which we consider as qualitatively \textit{optimistic, moderate, \& pessimistic} respectively.  Out of the $\sim 70$ pulsars included in the PTA public data releases, many have characterized red-noise processes and many have no observable signs of red noise.  This means that for our mock PTA, with \textit{uniform} noise characteristics in all pulsars, the white-noise only model `a' is likely \textit{overly}-optimistic, while model `e' is likely \textit{very}-pessimistic and model `d' may be somewhat pessimistic as well.  Unfortunately, only once each pulsar is observed over a duration comparable to the GW periods they are used to probe, will we have accurate measurements of their red-noise characteristics.

        The properties of the GWF thus depend not only on the distribution of the loudest sources, but also the frequency bin-width, and the distribution of all systems in that bin.  These effects are taken into account in the DS by including the background in the `noise' term, when calculating an SNR. A sample calculation of the strain, SNR, and detection probability (DP) for the GWF are shown in \figref{fig:gwf-det-stats}, for noise-model `d', at a number of sample frequencies.  While the single-source strains and resulting SNR are distributed around a well-defined peak, the distributions of DP end up being fairly flat because of their strong sensitivity to SNR.  These relatively flat DP distributions lead to large variations between realizations of the foreground.  The error bars included in our figures should thus be kept in mind.  The standard deviations in DP are fairly insensitive to increases in the number of realizations we consider, implying that the size of our underlying MBHB population may be the limiting factor (discussed further in \secref{sec:caveats}).

        To analyze the properties of the GWF, we define its sources as those which contribute at least a fraction $\forefac$ of the total GW energy in that bin, i.e.~$(\hcfore)^2 \gtrsim \forefac \, (\hc)^2 = \forefac [(\hcfore)^2 + (\hcback)^2]$.  In our analysis we explore different values of $\forefac$, but adopt a fiducial value of $\forefac = 0.5$, as used in \figref{fig:gwb_singles_zoom}.  In practice, the level at which a single source is discernible will depend on the overall source and PTA properties, especially via the signal-to-noise ratio (SNR), i.e.~$\forefac \propto \textrm{SNR}^{-1}$.  With a sufficient number of pulsars contributing comparatively to the SNR, a PTA can even discern multiple single-sources in a single frequency-bin \citep{babak2012, boyle2012, Petiteau2013}, making $\forefac = 0.5$ conservative in the eventual high-SNR regime.

\section{Results}
    \label{sec:results}

    \subsection{The Structure of the Low-Frequency GW Sky}

        \begin{figure}
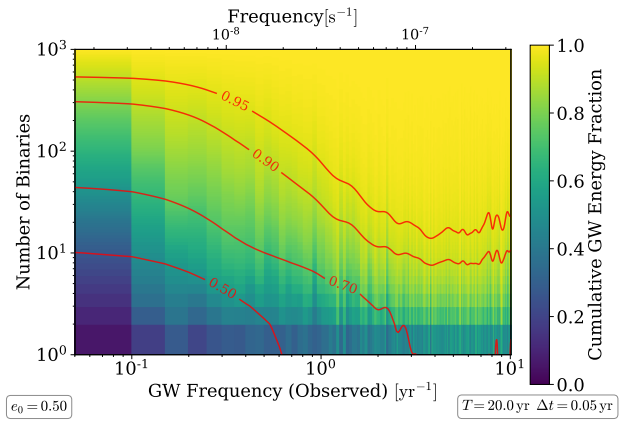

        \centering
        \includegraphics[width=\columnwidth]{{{fig2}}}
        \caption{The cumulative fraction of GW energy produced by the loudest binaries in each frequency bin.  The number of sources dominating the GW energy contribution drops rapidly with increasing frequency in correspondence with the overall number of binaries at the corresponding separations.  While the relative contribution from the loudest sources increases at high frequencies, the overall amplitude of GW signals simultaneously decreases.}
        \label{fig:num-loudest-ecc-evo-0.5}
        \end{figure}

        Figure~\ref{fig:num-loudest-ecc-evo-0.5} shows the median, cumulative fraction of GW energy contributed by the loudest binaries in each frequency bin.  The number of sources which contribute significantly to the GW energy density falls rapidly with increasing frequency in direct proportion to binary residence time.  For a circular binary hardening solely due to GW emission, the residence time scales with frequency as $\tgw \propto f^{-8/3}$.  At $\fgw \sim (10 \, \yr)^{-1}$, the median number of binaries producing $50\%$ of the GWB energy is $\sim 10$, while by $\fgw \sim 1 \, \pyr$, that number falls to $\sim 1$.  These results can be compared to those of \citet[][Fig.~2]{ravi2012} who find fairly consistent values, although more sources contributing at low frequencies, which implies a lower GWF rate.

        \begin{figure}
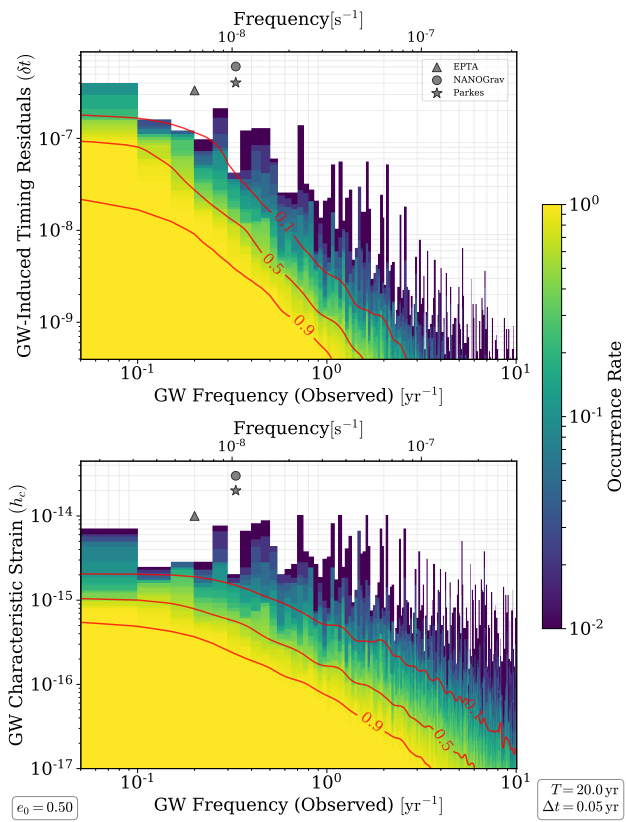

        \centering
        \includegraphics[width=\columnwidth]{{{fig3}}}
        \caption{The rate at which single-source timing residuals and strains of a given amplitude are produced in each frequency bin.  Below $f \sim \freqper{10}$, each bin typically has a source producing residuals of $\sim 10$ -- $100 \, \nanos$ ($\hc \sim 10^{-15}$).  The current single-source upper limits from PTA are also shown from the EPTA \citep{babak2015}, NANOGrav \citep{arzoumanian201404}, and PPTA \citep{zhu2014}.}
        \label{fig:loudest-strains-times_ecc-evo-0.5}
        \end{figure}

        Figure~\ref{fig:loudest-strains-times_ecc-evo-0.5} shows the probability of a single source producing timing residuals and strains above a given value.  At $\fgw \sim 1 \, \pyr$, $50\%$ of our realizations have a MBH binary with strain above $\sim 2\E{-16}$ or a timing residual of $\sim 1 \, \nanos$.  Our $10\%$ expectations are about an order of magnitude lower than recent PTA upper limits for foreground sources: $\sim 6\E{-15}$ -- $10^{-14}$ at $\sim (5 \, \yr)^{-1}$ by the EPTA \citep{babak2015}, $3\E{-14}$ at $\sim (3 \, yr)^{-1}$ by NANOGrav \citep{arzoumanian201404}, $2\E{-14}$ at $\sim (3 \, yr)^{-1}$ by the PPTA \citep{zhu2014}.  These strains and timing residuals are a few times higher than those of \citet{sesana2009} who predict at least one source with timing residuals between $\sim 5$ -- $50 \, \nanos$ after $T = 5\, \yr$.

        \begin{figure}
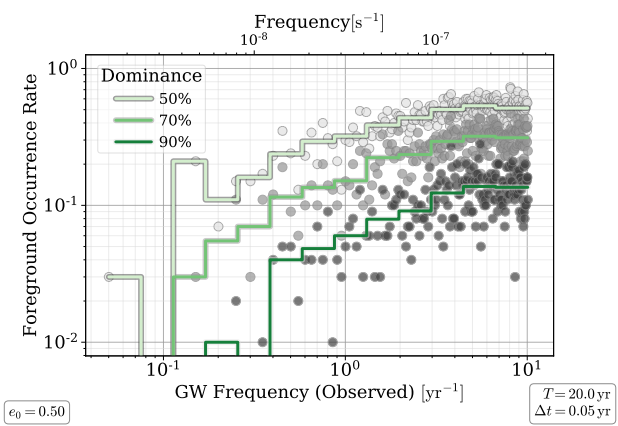

        \centering
        \includegraphics[width=\columnwidth]{{{fig4}}}
        \caption{The fraction of realizations with a foreground source in each frequency bin (circles) and averaged over frequencies (lines).  From light to dark, different criteria of foreground sources are show: ${\forefac = 0.5, 0.7 \, \& \, 0.9}$.  Single sources which are $\sim 10 \times$ louder than the background ($\forefac = 0.9$) are generally one-tenth as common as those equal to the background ($\forefac = 0.5$), at $f \sim \freqper{5}$ occurring $\sim 1\%$ and $\sim 10\%$ of the time respectively.}
        \label{fig:foreground-rate_ecc-evo-0.5}
        \end{figure}

        The previous figures examined the properties of all binaries.  Figure~\ref{fig:foreground-rate_ecc-evo-0.5} shows the fraction of realizations containing a foreground source in each frequency bin (circles) for three different foreground factors: ${\forefac = 0.5, 0.7 \, \& \, 0.9}$.  The average over fixed logarithmic frequency intervals is also shown (lines).  While the typical amplitude of foreground strains decreases significantly at higher frequencies, the ever decreasing number of sources contributing at those frequencies leads to higher GWF rates.

        \begin{figure}
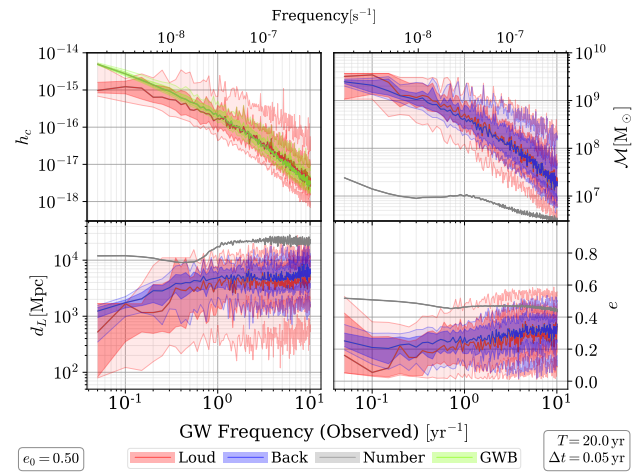

        \centering
        \includegraphics[width=\columnwidth]{{{fig5}}}
        \caption{Properties of low-frequency GW sources showing 1- \& 2- $\sigma$ contours over 100 realizations.  Properties of the loudest source in each frequency bin (red) are compared to those of all other systems (weighted by GW-energy; blue).  The median properties of all systems (unweighted) are also shown (grey).  Loud sources have much broader distributions of parameters, but tend to be slightly more massive, nearer, and lower eccentricity than the corresponding background sources when the latter are weighted by their GW energy.}
        \label{fig:gwf_properties}
        \end{figure}

        The parameters of the binaries producing low-frequency GW signals are shown in \figref{fig:gwf_properties}.  The properties of the loudest source in each frequency bin are plotted in red, and those of all other systems (weighted by GW energy) in blue.  For comparison, the median properties of all systems (unweighted) are shown in grey.  In the strain panel (upper-left), the GWB itself is shown for comparison in green as the typical strains of background sources are on the order of $10^{-19}$ -- $10^{-20}$.  The trends in binary parameters are driven by the convolution of hardening timescale and the number density of MBH binaries which falls strongly with total mass.  Higher-mass binaries harden faster, and all systems spend less and less time at smaller separations.  At low frequencies, where massive binaries are still numerous, they dominate the population and even occur at smaller distances.  Overall, loud sources have much broader distributions of parameters, but tend to be slightly more massive, nearer, and lower eccentricity than the corresponding background sources.

    \subsection{Parametric Dependencies}

        \begin{figure}
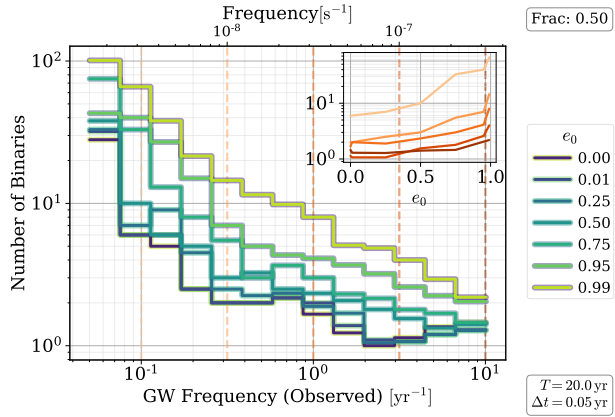

        \centering
        \includegraphics[width=\columnwidth]{{{fig6}}}
        \caption{Number of binaries contributing $50\%$ of the GW energy for each eccentricity model.  The inset shows the trends versus eccentricity at each of the orange highlighted frequencies.  As eccentricity increases, the number of contributing systems increases and the drop-off at higher frequencies becomes more gradual.}
        \label{fig:num-loudest_ecc}
        \end{figure}

        Eccentricity, by shifting the distribution of emitted GW energy versus frequency, effects the number of sources dominating GW signals.  Figure~\ref{fig:num-loudest_ecc} shows the number of loudest binaries contributing $0.5$ of the total GW energy for each eccentricity model.  The inset panel shows the trends versus eccentricity at the orange-highlighted frequencies.  As eccentricity increases, more sources contribute to the GW energy at all frequencies, but the effect is strongest at low to intermediate frequencies (\mbox{$\sim 0.1$ -- $0.3 \, \pyr$}).

        Varying the stellar scattering efficiency, unlike eccentricity, has very little effect on the number of binaries contributing significant GW energy.  While the effectiveness of scattering modulates the overall merger rate and total GW energy, it does not redistribute energy over frequency.  This fact is echoed in the amplitudes of the loudest sources which also show virtually no dependence on environmental hardening rate.  The latter is true, although to a lesser extent, with varying eccentricity.

        \begin{figure}
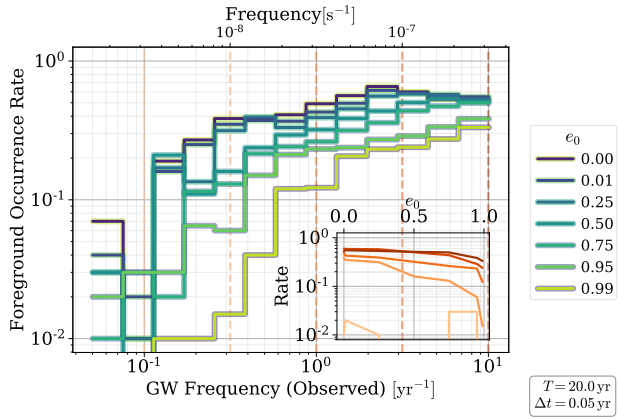

        \centering
        \includegraphics[width=\columnwidth]{{{fig7}}}
        \caption{The occurrence rate of foreground sources ($\forefac = 0.5$) per frequency bin for each initial eccentricity model.  For low-to-moderate initial eccentricities ($\eccinit \lesssim 0.5$) the rate plateaus to $\sim 50\%$ at frequencies $f \gtrsim 2 \, \pyr$, and at higher eccentricities ($\eccinit \gtrsim 0.75$) the occurrence rate drops rapidly at lower frequencies ($f \lesssim \freqper{2}$).  The inset panel shows the foreground occurrence rate versus eccentricity for the orange highlighted frequencies.}
        \label{fig:singles-frac_vs-ecc}
        \end{figure}

        Despite the weak effect of eccentricity on the loudness of sources, their varying number noticeably alters the rate of foreground source occurrences.  Figure~\ref{fig:singles-frac_vs-ecc} shows the rate at which a foreground source appears per frequency bin, for each eccentricity model.  Because more sources contribute significantly at higher eccentricities, the rate of foreground sources decreases.  At higher eccentricities, single sources have a higher GWB amplitude to compete with, at the same time as their own power is spread out over a broader frequency range.  Overall, initial eccentricities $\eccinit \gtrsim 0.9$ produce 2 -- 10 times fewer foreground sources as $\eccinit \sim 0.0$.

        Eccentricity also slightly increases the rate at which binaries harden, which has some effect on the GWB amplitude but very little effect on the GWF occurrence rate.  This can be seen in \figref{fig:singles-frac_vs-lc} which shows GWF rate for varying loss-cone refilling parameters ($\frefill$)---a good proxy for the overall degree of environmental coupling.  The increased hardening rate from $\frefill$ is substantially stronger than that of eccentricity (for these frequencies), and produces virtually no change to the GWF rate.

        \begin{figure}
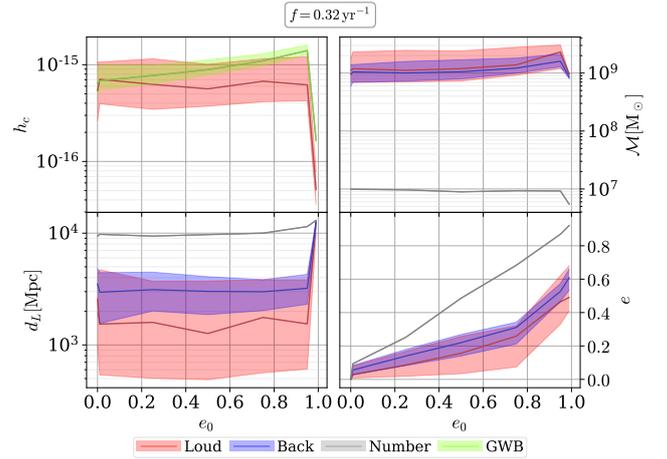

        \centering
        \includegraphics[width=\columnwidth]{{{fig8}}}
        \caption{Properties of the GW sources at $f = 0.32 \, \pyr = 10 \, \textrm{nHz}$ versus initial eccentricity.    Except for the highest eccentricity model ($\eccinit = 0.99$), the properties of the foreground are insensitive to evolutionary models.  There is a slight trend for increasing strains and chirp masses with both increasing initial-eccentricity and loss-cone efficiency.}
        \label{fig:gw-props_ecc}
        \end{figure}

        As alluded to previously, the binary parameters of sources contributing to the GWB and GWF are mostly insensitive to hardening models.  Figure~\ref{fig:gw-props_ecc} shows the properties of the loudest sources in each bin (red) compared to those of the background, weighted by GW energy, (blue) as a function of initial-eccentricity, at a frequency of $f = 0.32 \, \pyr$.  The distribution of unweighted properties are also plotted (grey) for comparison.  In each case, a one-sigma interval is shown.  In the most extreme eccentricity model, $\eccinit = 0.99$, the distances to both loud and background sources increases drastically and the strain drops correspondingly.  In this case, the eccentricity is extreme enough that the majority of the GW energy in this regime is shifted out of band.  Otherwise typical properties are constant as eccentricity varies, as is the case with varying stellar scattering efficiencies.

        The lower right panel in \figref{fig:gw-props_ecc} shows the eccentricity distributions of sources at $f = 0.32 \, \pyr$.  The eccentricity in the unweighted population of all binaries is nearly linear with initial eccentricity.  Loud sources, and the binaries which contribute most to the background, however, have significantly damped eccentricities which are much lower in the PTA band than their initial values\footnote{\citet{paper2} shows in detail the evolution of eccentricity as binaries harden}.  The loudest sources tend to have lower median eccentricities by $\sim 0.1$, and the low-end of their $68\%$ interval is typically half that of background sources.  More massive systems circularize more rapidly, but the tendency for lower eccentricities in single sources is likely a selection bias as they are the systems which emit GW energy in a more concentrated frequency range (i.e.~nearer the $n=2$ harmonic).

    \subsection{Times to Detection}

        \begin{figure}
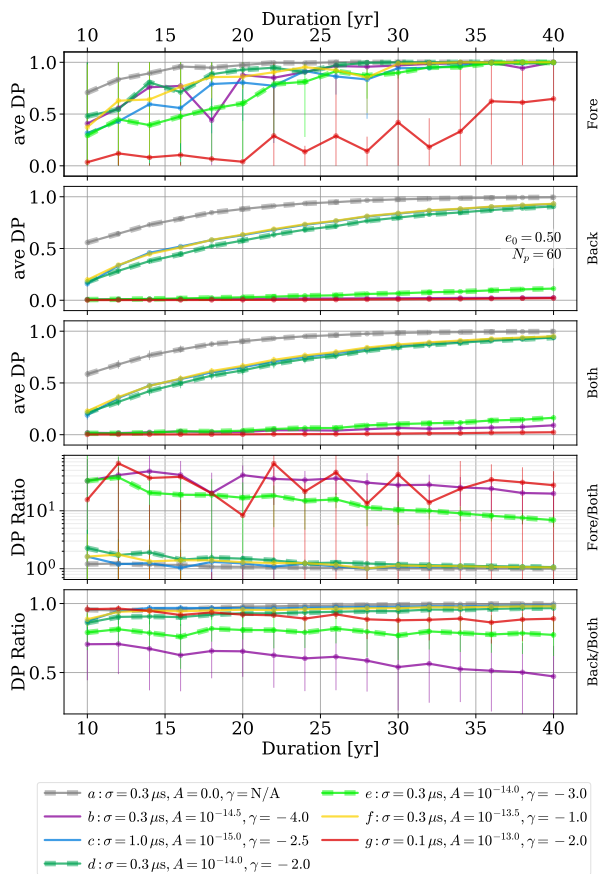

        \centering
        \includegraphics[width=\columnwidth]{{{fig9}}}
        \caption{Detection Probability (DP) versus observing duration for mock PTA with a variety of noise models (lines).  From top-to-bottom, the panels show DP for the `Fore'ground, `Back'ground, and the combination of `both' background and loudest sources (still measured using the background detection statistics).  The bottom two panels show the ratio of detection probabilities for the foreground versus both, and the background versus both.  Noise models `a', `d' \& 'e', which we focus on, are highlighted with dashes.  The DP for the foreground is effectively always higher than that of the background (or combination).  Note that for the `back' and `both' signals, the noise models `c' \& `f' nearly-perfectly overlap.}
        \label{fig:dp-dur_noise}
        \end{figure}

        In \figref{fig:dp-dur_noise}, we show detection probability (DP) versus time for the $\eccinit = 0.5$ model, and a PTA with 60 pulsars for each noise model from \tabref{tab:noise} and \figref{fig:noise}.  For comparison, the IPTA first data release included almost 50 pulsars, with a median observing duration of $T \sim 11 \, yr$.  Forecasts for expansions typically assume 6 new pulsars added to the IPTA per year \citep[e.g.][]{taylor2016}, but new pulsars, with very short observing baselines, will contribute very little to DP initially.  Recall, additionally, that we assume uniform noise characteristics for each pulsar in our mock arrays, while real PTA have pulsars with highly heterogenous noise characteristics which further complicates a direct comparison to our results.

        The first three panels of \figref{fig:dp-dur_noise} show DP for the foreground, background and combination.  Keeping the above caveats in mind, if we associate this mock PTA at 10 years with the current IPTA, then depending on noise model the expected DP would be anywhere between $\sim 0.0$ for severe red noise, to $\sim 0.7$ or $\sim 0.5$ for the foreground and background respectively for white-noise only.  Clearly, better understanding the red noise of observed pulsars is crucial to reliably forecasting low-frequency GW detections.

        The last two panels show the DP ratios of foreground-to-both, and background-to-both.  In all models, the GWF have uniformly higher DP than either the `back' or `both' signals.  In the case of no red-noise (`a', grey), the `back' and `both' DP are only slightly below that of the foreground.  Stronger noise, especially red noise, affects the GWB detection much more strongly than the GWF.  This is not surprising as red noise, by definition, affects the lowest frequency bins most strongly, which is where the GWB is loudest, but generally below the GWF peak (see, e.g.~\figref{fig:loudest-strains-times_ecc-evo-0.5}).

        With the GWB detection statistics, the noise models become highly stratified.  The shallow red-noise models---`c', `d', \& `f'---all perform comparably with $\sim 50\%$ lower DP than the white-noise only model at 10--15 yr.  At later times the shallow models approach the white-noise only values.  The steep red-noise models---`b', `e' \& `g'---also all perform comparably, but with near-zero DP up to $\sim 30 \, \yr$.  The `both' and `back' DP typically differ by at most $50\%$, while `fore' and `both' differ by over an order of magnitude in the steep red-noise models.  This suggests that the difference between GWF and GWB DP is driven largely by the nature of the detection statistic instead of simply the amount of GW power being analyzed.

        To observe the GWB, the correlations between pulsars are required to distinguish the GW signal from that of noise.  Because GWF sources will behave deterministically, they should be easier to distinguish.  At the same time, line-like noise sources (for example due to uncertainties in planetary ephemerides), or GWF sources with few resolved periods (thus more closely resembling red noise), may complicate the identification process via single pulsars.  If, instead, a correlated search is required, the recovery efficiency could become significantly lower --- perhaps yielding DP which lie below that of the GWB.

        \begin{figure}
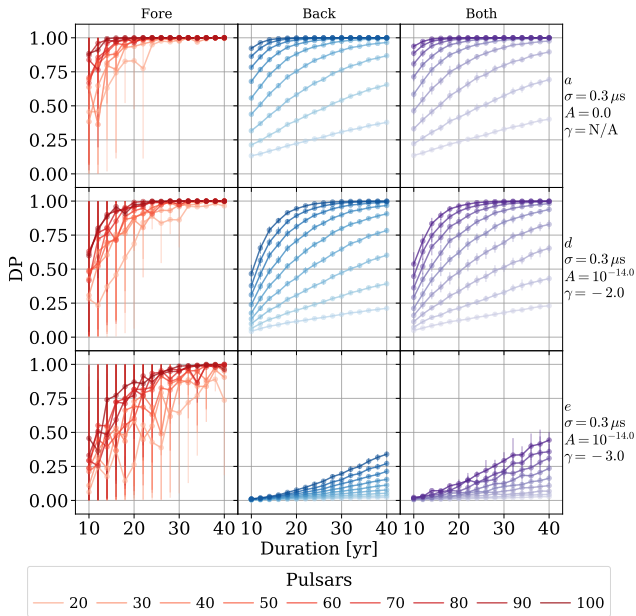

        \centering
        \includegraphics[width=\columnwidth]{{{fig10}}}
        \caption{Detection probability versus time for the foreground, background, and combination (`both'; columns), showing the representative noise-models `a', `d' and `e' (rows).  PTA with varying numbers of pulsars are illustrated with differing line colors.  The number of pulsars very strongly effects the DP of the GWB (`back' and `both') because they rely on cross-correlations to make detections, while the GWF is much less sensitive to pulsar number.  The GWF DP is also much more robust against worsening red noise as the foreground spectrum is flatter than that of the GWB.}
        \label{fig:dp-dur_pulsars}
        \end{figure}

        Due to the similarity between numerous noise models, our additional analysis focuses on the `a', `d' \& `e' configurations.  These can be considered qualitatively as optimistic, moderate \& pessimistic respectively, but keep in mind that realistic PTA pulsars have heterogenous noise properties possibly making `a' and `d'/`e' \textit{overly} optimistic \& pessimistic respectively (see \secref{sec:meth_ds}).  Figure~\ref{fig:dp-dur_pulsars} compares DP progressions for PTA with differing numbers of pulsars and these select noise models.  Because the GWB detection statistic depends on a cross-correlation between pulsars---i.e.~pulsar-pairs, it is far more sensitive to the number of pulsars included in the array.  Considering the `d' noise-model: at $T = 20 \, \yr$, the difference in DP between 20 and 80 pulsars is $70$ vs.~$95\%$ for the GWF, but $10$ vs.~$80\%$ for the GWB.  In the highly pessimistic noise-model `e', $50\%$ DP isn't reached for the background within $40 \, \yr$, even for 100 pulsars, while the foreground reaches $50\%$ DP in $\sim 15 \, \yr$.

        \begin{figure}
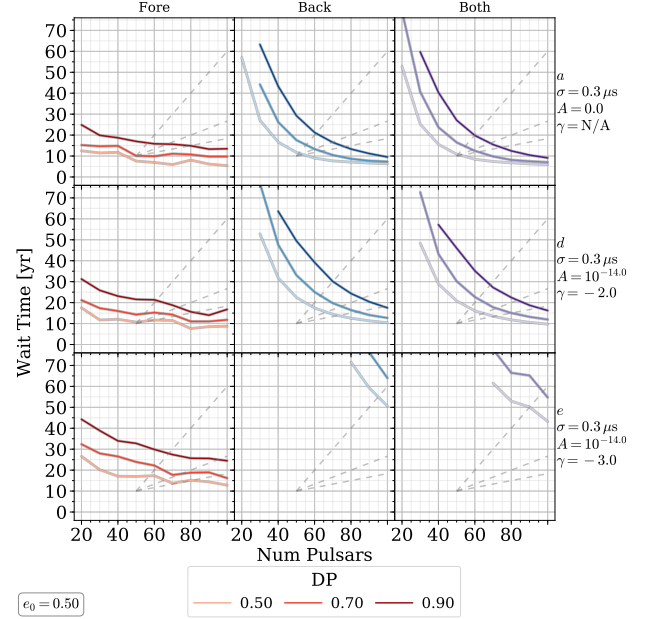

        \centering
        \includegraphics[width=\columnwidth]{{{fig11}}}
        \caption{Time to reach a given detection probability versus number of PTA pulsars.  Optimistic `a' (white noise only), moderate `d', and pessimistic `e' noise models are shown.  The grey dashed lines show sample expansion rates for a PTA starting with 50 pulsars after 10 years of observations and adding 1, 3 \& 6 pulsars per year.  If we consider the 3/yr expansion rate with the `d' noise model realistic for the IPTA, then we would expect to reach $90\%$ DP in roughly $8$ \& $12 \, \yr$ (overall observing baselines of $18$ and $22 \, \yr$) for the GWF \& GWB respectively.}
        \label{fig:wait-times_pulsars}
        \end{figure}

        A comparison between the DP curves for varying eccentricity models and varying loss-cone parameters are shown in Figs.~\ref{fig:dp-dur_ecc}~\&~\ref{fig:dp-dur_lc}.  In general the DP varies by $\sim 15$ -- $25\%$ based on varying evolutionary parameters.  In the case of the highest eccentricities, $\eccinit = 0.95$ \& $0.99$, the GWB DP drops drastically, and in the latter case is effectively unobservable for all of the noise and PTA models considered.  For low to moderate eccentricities ($\eccinit \lesssim 0.75$) the time to reach a given DP typically varies by $\sim 5 \, \yr$, while for varying environmental coupling it varies by up to $\sim 10 \, \yr$.  Still, the dependence on noise model and number of pulsars tends to be more significant than between evolutionary parameters.

        The time to reach a given DP is plotted versus number of pulsars in \figref{fig:wait-times_pulsars}.  It typically takes $\sim 5$ -- $10 \, \yr$ to increase from a $50\%$ DP to $90\%$, but for low numbers of pulsars it can be as long as $\sim 20 \, \yr$.  As already discussed, GWB detection relies on correlations between pulsars which introduces a very strong dependence on the number of pulsars, which is clearly apparent.  In the modern `d' noise model, doubling the number of pulsars from 40 to 80 decreases the time to detection by almost a factor of 3: from 65 to 22 years (for DP $ = 0.9$).  GWF detection is much less sensitive to number of pulsars, with the time to detection decreasing from $\sim 23$ to $15 \, \yr$ from the same increase in pulsar number.

        Recall that the IPTA, for example, is expected to expand by roughly 6 pulsars per year, meaning that as it continues to collect data it not only moves upward in this figure, but also to the right.  The grey dashed lines assume a starting point of 50 pulsars after 10 years of observations (based on the first IPTA data release) with the addition of 1, 3 \& 6 pulsars per year shown in each line.  If the 3 pulsars/year line accurately reflects the IPTA expansion, after taking into account the decreased leverage of newly added pulsars, then we would expect the GWF \& GWB to reach $90\%$ DP after roughly $18$ \& $22 \yr$ respectively of total observing time (i.e.~$8$ \& $12 \, \yr$ of additional observations) for the moderate noise model `d'.

\section{Discussion}
    \label{sec:discussion}

    \subsection{Caveats}
        \label{sec:caveats}

        Some caveats should be borne in mind when examining our results.  First, the variations in detection probability between realizations of our foreground models are significant.  The foreground sources which end up being `detected' in our analysis are sampled from a representative distribution of binary parameters instead of being dominated by a handful of systems.  Still, the overall population of MBHB from Illustris may be insufficient to properly sample the full light-cone of observations.  Similarly, while we resample our populations to extrapolate to the volume of the universe and include Poisson noise, we are still subject to the same intrinsic systematics of our underlying source population.  This is all the more true in a GWF analysis which is by definition far more sensitive to individual sources than the GWB is.

        It is also important to note that our mock PTA models are suboptimal because they use uniform pulsar parameters across the array, as required by the detection statistics we use.  The time sampling and noise models we have used in our analysis are representative of the published specifications of PTA pulsars.  Still, real PTA include highly heterogenous populations of pulsars with varying sampling and noise characteristics.  For example, our pessimistic noise model (`e') is consistent with that of some pulsars, but in a heterogenous PTA, pulsars with such strong noise would contribute far less to the SNR than other pulsars with better noise levels.  The pessimistic model `e' is thus certainly \textit{overly} pessimistic.  While many pulsars have no observed red noise, our optimistic model `a'---with only white noise---is, similarly, likely \textit{overly} optimistic\footnote{At least when compared to a heterogenous PTA with the same total number of pulsars.}.  In addition to uncertainties in the most representative noise model, even choosing an accurate number of pulsars is non-trivial.  PTA continue to expand by adding new systems, which are a very important part of detection forecasts \citep[e.g.][]{taylor2016}, but have different frequency sensitivities and thus different leverages on SNR.

        Finally, the GWF detection statistics we use are suboptimal in at least two respects.  First, the statistics in \citet{rsg15} do not account for eccentric effects, and we treat the GW energy from single sources that is split across multiple frequency bins as entirely independent.  Secondly, the excess power statistics may not perform as well on data where noise processes are harder to distinguish from single GW sources.  The presence of line-like noise, due to uncertainties in planetary ephemerides (and their harmonics) for example; or instrumental effects, could introduce significant interference.  At the very least, it is likely that foreground sources with periods comparable to the observing duration will be hard to distinguish from red noise when using single pulsar searches.  This may necessitate correlated searches with lower recovery efficiencies.

    \subsection{Conclusions}

        Single GW sources resolvable above the background, which we refer to as \textbf{GW Foreground (GWF) sources, tend to be most detectable at frequencies near $\sim 0.1$ -- $0.3 \, \pyr$}.  At higher frequencies there are fewer sources which also tend to be less loud.  At lower frequencies the gravitational wave background (GWB) from all other sources is more likely to drown out the singles.  According to most of our models, we would expect there to be a foreground source with a characteristic strain of about $10^{-15}$ (or timing residual of $\sim 30 \, \nanos$) after $10 \, \yr$ of observations.  These values are roughly a factor of two higher than those predicted by \citet[][e.g.~Fig.~3]{sesana2009} and \citet[][e.g.~Fig.~4]{ravi2014}

        When taking eccentricity into account, the primary effect of non-circular evolution is to shift GW energy from lower to higher frequencies.  Thus, \textbf{increasing eccentricity decreases the occurrence rate of GWF sources}, especially at lower frequencies, both by increasing the GWB amplitude and by diffusing the single source strains themselves.  Changes in the effectiveness of stellar scattering, and more generally \textbf{the rate of environmental hardening, has a small effect on the properties and occurrence rates of the GWF}.  Measurements of the number of GWF sources can thus provide strong constraints on the eccentricity distribution of the underlying MBHB population.  Even in the absence of foreground detections, measuring the number of sources contributing to the background could provide the same information.  This could be done either by directly resolving numerous loud sources in particular frequency bins (i.e.~in frequency space), or indirectly by inference from the degree of anisotropy of the GWB (i.e.~in angular space).

        \textbf{Detection probabilities are usually higher for the GWF than for the GWB} --- indicating that the background may not be detected first, as has generally been expected.  We emphasize, however, that there is a large variance between realizations in our simulations suggesting that \textit{our population of MBHB may not be large enough for fully converged results}.  Based on our models, however, mock PTA comparable to the IPTA are able to reach high detection probabilities in $18$ \& $22 \, \yr$ of total observing time for the GWF \& GWB respectively, with moderate parameter assumptions\footnote{For comparison, median observing baselines for the IPTA are currently $\sim 10 \, \yr$.}.  Our detection probabilities for the GWB are closely in line with those of \citet{rsg15}, while our GWF values are notably higher for PTA with $\sim 50$ pulsars, though again consistent at $\sim 100$ pulsars.  The higher GWF values may be due to our higher single source strains, or possibly a tendency for our sources to reside at lower redshifts owing to our dynamical binary evolution which is not included in the previous, semi-analytic models.

        \textbf{Pulsar red-noise models have a very strong effect on detection probabilities and times to detection}, especially for the GWB.  Comparing between white-noise only, and a moderate red-noise model, the time-to-detection can increase by $50$ -- $100\%$.  In the case of uniformly severe red noise, prospects for detecting the GWB can become bleak, but many of the currently monitored PTA pulsars show no signs of red noise at all.  That being said, \textbf{the GWF is much less sensitive to red noise} as single sources are best detected at intermediate frequencies, unlike the GWB which, in our models, is almost always strongest at the lowest accessible frequency bins.  Varying eccentricity and environmental coupling have moderate effects on times to detection.  For our moderate noise model, \textbf{nearly-circular evolution takes $\sim 5\, \yr$ longer to detect than moderately-high eccentric evolution.  Similarly, best-case stellar scattering takes $\sim 10 \, \yr$ less time to detect than the worst case.}

        Red-noise models must be included when constructing realistic forecasts for PTA detection prospects, which has not been done in existing studies.  Furthermore, we hope that the red-noise characteristics of PTA pulsars will be more thoroughly explored in the context of the IPTA to better calibrate our expectations.  The development of more flexible detection statistics for low-frequency GW single sources would also be very helpful in constructing realistic PTA models and testing them with cosmological MBHB populations.

        Many studies have shown that the GWB may be within a decade or so of detection, which is consistent with the results presented here.  For the first time, however, we find that the GWF might be just as detectable or possibly even more so.  Similarly, PTA upper limits on GWF sources should also be used to constrain the MBHB population as is being done with GWB upper-limits.  To this end, additional studies, with larger populations of MBHB, should be explored.  Prospects for PTA detection of low-frequency GW seem very promising, and even with only upper limits, we stand at the precipice of making substantial progress in our understanding of MBH binaries and their evolution.


\vspace{-0.15in}
\section*{Acknowledgments}
    Many of the computations in this paper were run on the Odyssey cluster supported by the FAS Division of Science, Research Computing Group at Harvard University.  We greatly appreciate the help and generous support provided by the Research Computing Group and by the Computational Facility at the Harvard-Smithsonian Center for Astrophysics.  This research made use of \astropy, a community-developed core Python package for Astronomy \citep{astropy2013}, in addition to \scipy~\citep{scipy}, \ipython~\citep{ipython}, \& \numpy~\citep{numpy2011}.  All figures were generated using \matplotlib~\citep{matplotlib2007}.

\let\oldUrl\url
\renewcommand{\url}[1]{\href{#1}{Link}}

\quad{}
\bibliographystyle{mnras}
\bibliography{refs}

\appendix


    \section{Additional Figures}

        \begin{figure*}
        \centering
        \includegraphics[width=0.8\textwidth]{{{fig_a1}}}
        \caption{Noise parameters for all pulsars with measured red-noise characteristics in the PTA public data releases \citep{desvignes2016, Verbiest201602, arzoumanian2015a, reardon2016, caballero2016, lentati2016}.  When multiple PTA have independent red-noise fits we simply include each characterization independently as they often differ substantially.  The first two columns show the white- and red- noise amplitude in units of dimensionless strain at $f = \freqper{10}$.  The third and fourth columns show the red-noise amplitude and spectral index (see: \refeq{eq:red-noise}).}
        \label{fig:noise}
        \end{figure*}

        \begin{figure}
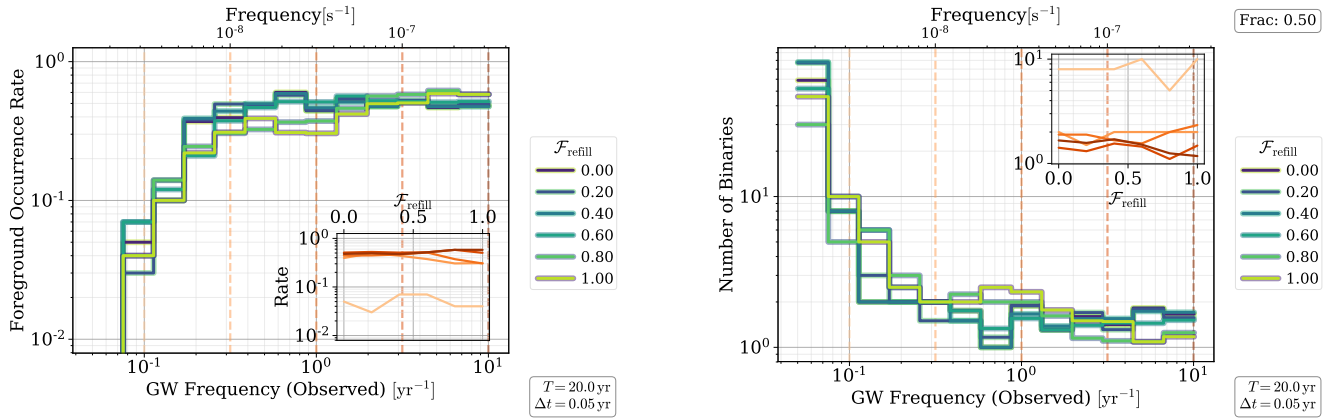

        \centering
        \includegraphics[width=\columnwidth]{{{fig_a2}}}
        \caption{The occurrence rate of foreground sources per frequency bin for each loss-cone efficiency model, using $\forefac = 0.5$.  The inset panel shows the foreground occurrence rate versus eccentricity for the highlighted frequencies (dashed orange lines).  While the occurrence rate of foreground sources drops significantly at frequencies below $\sim \freqper{5}$, it remains relatively constant at higher frequencies, unlike in eccentric models.  There is also almost no dependence of the foreground rate with stellar scattering efficiency ($\frefill$).}
        \label{fig:singles-frac_vs-lc}
        \end{figure}

        \begin{figure}
        \centering
        \includegraphics[width=\columnwidth]{{{fig_a3}}}
        \caption{Number of binaries contributing $50\%$ of the GW energy for each stellar scattering model.  The inset panel shows trends versus scattering efficiency at each of the highlighted frequencies (dashed orange).  Overall, there is no significant dependence on the relative contribution of binaries from varying stellar scattering.}
        \label{fig:num-loudest_lc}
        \end{figure}

        \begin{figure}
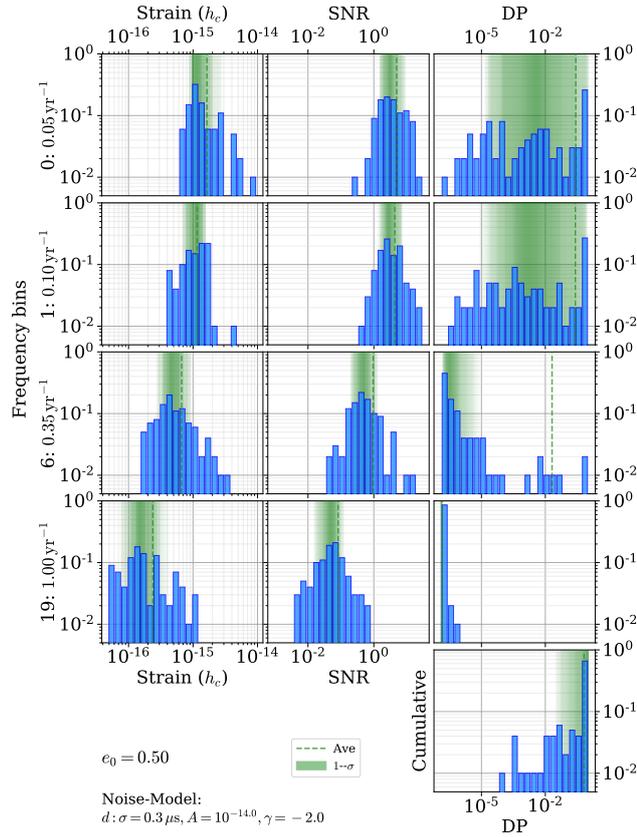

        \centering
        \includegraphics[width=\columnwidth]{{{fig_a4}}}
        \caption{Detection statistics for the GWF at different frequencies.  The $\eccinit = 0.5$ model is shown, observed by a PTA with 60 pulsars and noise-model `d'.  Distributions calculated over 200 realizations are shown in blue, with $1 \, \sigma$ contours in green, and the average values marked by the dashed green line.  The strain from each frequency bin, along with the GWB strain and the pulsar noise characteristics, determines the SNR which then maps to a detection probability.  The cumulative DP of detecting at least one single-source, over all frequency bins, is shown in the lower right.}
        \label{fig:gwf-det-stats}
        \end{figure}

        \begin{figure}
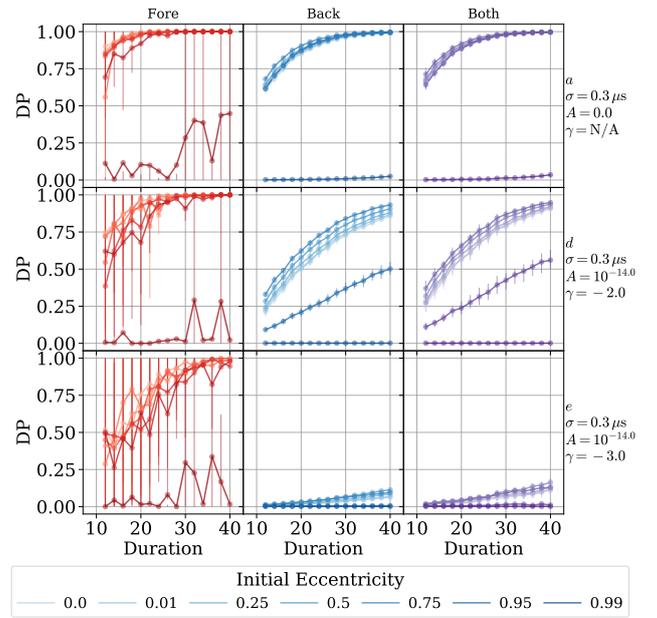

        \centering
        \includegraphics[width=\columnwidth]{{{fig_a5}}}
        \caption{Detection probability is shown versus time for a PTA with 60 pulsars for each eccentricity model.  There is mostly a moderate dependence of DP on eccentricity model, causing a variation of $\sim 5\,\yr$ to reach a given DP.  For the most extreme, $\eccinit = 0.99$, case for the GWF, and additionally the $\eccinit = 0.95$ case for the GWB, the DP drops drastically.  Overall, the red-noise model still has a larger effect on DP.}
        \label{fig:dp-dur_ecc}
        \end{figure}

        \begin{figure}
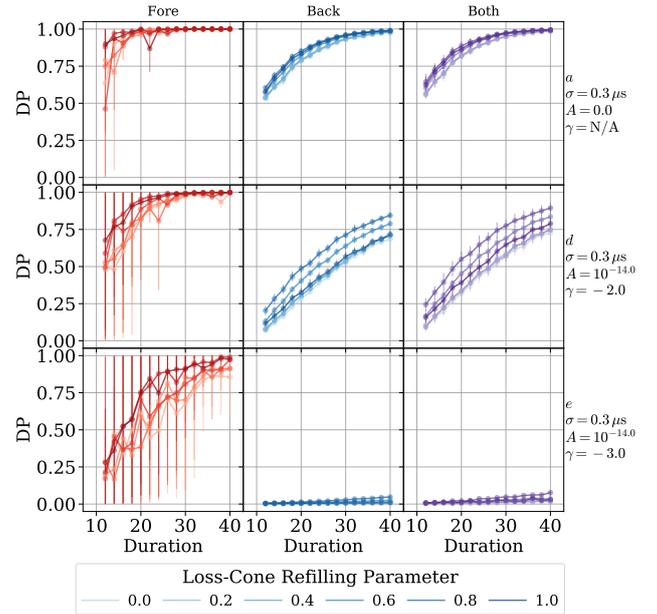

        \centering
        \includegraphics[width=\columnwidth]{{{fig_a6}}}
        \caption{Detection probability is shown versus time for a PTA with 60 pulsars for each stellar scattering model.  There is a moderate dependence of DP on scattering efficiency, causing a variation of $\sim 5$ -- $10\,\yr$ to reach a given DP.  Overall, the red-noise model still has a larger effect on DP.}
        \label{fig:dp-dur_lc}
        \end{figure}

\label{lastpage}

\end{document}

%% file: AstroComs.tex
\newcommand{\tr}[1]{\textrm{#1}}
\newcommand{\trt}[1]{\textrm{\tiny{#1}}}

\newcommand{\msol}{\tr{M}_{\odot}}

\newcommand{\lr}[1]{\left(#1\right)}
\newcommand{\scale}[3][]{\lr{ \frac{#2}{#3} }^{#1}}

\newcommand{\E}[1]{\times\nobreak10^{#1}}

\newcommand{\secref}[1]{\textsection\ref{#1}}
\newcommand{\figref}[1]{Fig.~\ref{#1}}
\newcommand{\refeq}[1]{{Eq.~\ref{#1}}}      
\newcommand{\tabref}[1]{{Table~\ref{#1}}}

\newcommand{\ayr}{A_{\trt{yr}^{-1}}}        
\newcommand{\yr}{\,\mathrm{yr}}        

\newcommand{\pyr}{\textrm{yr}^{-1}}

\newcommand{\comdist}{d_\trt{c}}   
\newcommand{\astropy}{\texttt{Astropy}}
\newcommand{\matplotlib}{\texttt{matplotlib}}
\newcommand{\numpy}{\texttt{NumPy}}
\newcommand{\scipy}{\texttt{SciPy}}
\newcommand{\ipython}{\texttt{ipython}}

%% file: JAbs.tex
\def\apj{Astrophys. J.}       		        	 	
\def\apjl{Astrophys. J. Lett.}             		

\def\pasa{PASA}
\def\mnras{Mon. Not. R. Astron. Soc.}        
\def\prd{Phys. Rev. D}      				
\def\aap{Astron. Astrophys.}               		

%% file: ms.bbl
\begin{thebibliography}{}
\makeatletter
\relax
\def\mn@urlcharsother{\let\do\@makeother \do\$\do\&\do\#\do\^\do\_\do\%\do\~}
\def\mn@doi{\begingroup\mn@urlcharsother \@ifnextchar [ {\mn@doi@}
  {\mn@doi@[]}}
\def\mn@doi@[#1]#2{\def\@tempa{#1}\ifx\@tempa\@empty \href
  {http://dx.doi.org/#2} {doi:#2}\else \href {http://dx.doi.org/#2} {#1}\fi
  \endgroup}
\def\mn@eprint#1#2{\mn@eprint@#1:#2::\@nil}
\def\mn@eprint@arXiv#1{\href {http://arxiv.org/abs/#1} {{\tt arXiv:#1}}}
\def\mn@eprint@dblp#1{\href {http://dblp.uni-trier.de/rec/bibtex/#1.xml}
  {dblp:#1}}
\def\mn@eprint@#1:#2:#3:#4\@nil{\def\@tempa {#1}\def\@tempb {#2}\def\@tempc
  {#3}\ifx \@tempc \@empty \let \@tempc \@tempb \let \@tempb \@tempa \fi \ifx
  \@tempb \@empty \def\@tempb {arXiv}\fi \@ifundefined
  {mn@eprint@\@tempb}{\@tempb:\@tempc}{\expandafter \expandafter \csname
  mn@eprint@\@tempb\endcsname \expandafter{\@tempc}}}

\bibitem[\protect\citeauthoryear{{Arzoumanian} et~al.,}{{Arzoumanian}
  et~al.}{2014}]{arzoumanian201404}
{Arzoumanian} Z.,  et~al., 2014, \mn@doi [\apj] {10.1088/0004-637X/794/2/141},
  \href {http://adsabs.harvard.edu/abs/2014ApJ...794..141A} {794, 141}

\bibitem[\protect\citeauthoryear{{Arzoumanian} et~al.,}{{Arzoumanian}
  et~al.}{2016}]{Arzoumanian201508}
{Arzoumanian} Z.,  et~al., 2016, \mn@doi [\apj] {10.3847/0004-637X/821/1/13},
  \href {http://adsabs.harvard.edu/abs/2016ApJ...821...13A} {821, 13}

\bibitem[\protect\citeauthoryear{{Astropy Collaboration} et~al.,}{{Astropy
  Collaboration} et~al.}{2013}]{astropy2013}
{Astropy Collaboration} et~al., 2013, \mn@doi [\aap]
  {10.1051/0004-6361/201322068}, \href
  {http://adsabs.harvard.edu/abs/2013A%26A...558A..33A} {558, A33}

\bibitem[\protect\citeauthoryear{{Babak} \& {Sesana}}{{Babak} \&
  {Sesana}}{2012}]{babak2012}
{Babak} S.,  {Sesana} A.,  2012, \mn@doi [\prd] {10.1103/PhysRevD.85.044034},
  \href {http://adsabs.harvard.edu/abs/2012PhRvD..85d4034B} {85, 044034}

\bibitem[\protect\citeauthoryear{{Babak} et~al.,}{{Babak}
  et~al.}{2016}]{babak2015}
{Babak} S.,  et~al., 2016, \mn@doi [\mnras] {10.1093/mnras/stv2092}, \href
  {http://adsabs.harvard.edu/abs/2016MNRAS.455.1665B} {455, 1665}

\bibitem[\protect\citeauthoryear{{Bertotti}, {Carr}  \& {Rees}}{{Bertotti}
  et~al.}{1983}]{bertotti1983}
{Bertotti} B.,  {Carr} B.~J.,   {Rees} M.~J.,  1983, \mn@doi [\mnras]
  {10.1093/mnras/203.4.945}, \href
  {http://adsabs.harvard.edu/abs/1983MNRAS.203..945B} {203, 945}

\bibitem[\protect\citeauthoryear{{Blandford}, {Romani}  \&
  {Narayan}}{{Blandford} et~al.}{1984}]{blandford1984}
{Blandford} R.,  {Romani} R.~W.,   {Narayan} R.,  1984, \mn@doi [Journal of
  Astrophysics and Astronomy] {10.1007/BF02714466}, \href
  {http://adsabs.harvard.edu/abs/1984JApA....5..369B} {5, 369}

\bibitem[\protect\citeauthoryear{{Boyle} \& {Pen}}{{Boyle} \&
  {Pen}}{2012}]{boyle2012}
{Boyle} L.,  {Pen} U.-L.,  2012, \mn@doi [\prd] {10.1103/PhysRevD.86.124028},
  \href {http://adsabs.harvard.edu/abs/2012PhRvD..86l4028B} {86, 124028}

\bibitem[\protect\citeauthoryear{{Caballero} et~al.,}{{Caballero}
  et~al.}{2016}]{caballero2016}
{Caballero} R.~N.,  et~al., 2016, \mn@doi [\mnras] {10.1093/mnras/stw179},
  \href {http://adsabs.harvard.edu/abs/2016MNRAS.457.4421C} {457, 4421}

\bibitem[\protect\citeauthoryear{{Charisi}, {Bartos}, {Haiman}, {Price-Whelan},
  {Graham}, {Bellm}, {Laher}  \& {M{\'a}rka}}{{Charisi}
  et~al.}{2016}]{charisi2016}
{Charisi} M.,  {Bartos} I.,  {Haiman} Z.,  {Price-Whelan} A.~M.,  {Graham}
  M.~J.,  {Bellm} E.~C.,  {Laher} R.~R.,   {M{\'a}rka} S.,  2016, \mn@doi
  [\mnras] {10.1093/mnras/stw1838}, \href
  {http://adsabs.harvard.edu/abs/2016MNRAS.463.2145C} {463, 2145}

\bibitem[\protect\citeauthoryear{{Corbin} \& {Cornish}}{{Corbin} \&
  {Cornish}}{2010}]{corbin2010}
{Corbin} V.,  {Cornish} N.~J.,  2010, preprint, \href
  {http://adsabs.harvard.edu/abs/2010arXiv1008.1782C} {} (\mn@eprint {arXiv}
  {1008.1782})

\bibitem[\protect\citeauthoryear{{Demorest} et~al.,}{{Demorest}
  et~al.}{2013}]{demorest2013}
{Demorest} P.~B.,  et~al., 2013, \mn@doi [\apj] {10.1088/0004-637X/762/2/94},
  \href {http://adsabs.harvard.edu/abs/2013ApJ...762...94D} {762, 94}

\bibitem[\protect\citeauthoryear{{Desvignes} et~al.,}{{Desvignes}
  et~al.}{2016}]{desvignes2016}
{Desvignes} G.,  et~al., 2016, \mn@doi [\mnras] {10.1093/mnras/stw483}, \href
  {http://adsabs.harvard.edu/abs/2016MNRAS.458.3341D} {458, 3341}

\bibitem[\protect\citeauthoryear{{Detweiler}}{{Detweiler}}{1979}]{detweiler1979}
{Detweiler} S.,  1979, \mn@doi [\apj] {10.1086/157593}, \href
  {http://adsabs.harvard.edu/abs/1979ApJ...234.1100D} {234, 1100}

\bibitem[\protect\citeauthoryear{{Ellis}}{{Ellis}}{2013}]{ellis2013}
{Ellis} J.~A.,  2013, \mn@doi [Classical and Quantum Gravity]
  {10.1088/0264-9381/30/22/224004}, \href
  {http://adsabs.harvard.edu/abs/2013CQGra..30v4004E} {30, 224004}

\bibitem[\protect\citeauthoryear{{Ellis}, {Siemens}  \& {Creighton}}{{Ellis}
  et~al.}{2012}]{ellis201204}
{Ellis} J.~A.,  {Siemens} X.,   {Creighton} J.~D.~E.,  2012, \mn@doi [\apj]
  {10.1088/0004-637X/756/2/175}, \href
  {http://adsabs.harvard.edu/abs/2012ApJ...756..175E} {756, 175}

\bibitem[\protect\citeauthoryear{{Enoki}, {Inoue}, {Nagashima}  \&
  {Sugiyama}}{{Enoki} et~al.}{2004}]{enoki2004}
{Enoki} M.,  {Inoue} K.~T.,  {Nagashima} M.,   {Sugiyama} N.,  2004, \mn@doi
  [\apj] {10.1086/424475}, \href
  {http://adsabs.harvard.edu/abs/2004ApJ...615...19E} {615, 19}

\bibitem[\protect\citeauthoryear{{Estabrook} \& {Wahlquist}}{{Estabrook} \&
  {Wahlquist}}{1975}]{estabrook1975}
{Estabrook} F.~B.,  {Wahlquist} H.~D.,  1975, \mn@doi [General Relativity and
  Gravitation] {10.1007/BF00762449}, \href
  {http://adsabs.harvard.edu/abs/1975GReGr...6..439E} {6, 439}

\bibitem[\protect\citeauthoryear{{Foster} \& {Backer}}{{Foster} \&
  {Backer}}{1990}]{foster1990}
{Foster} R.~S.,  {Backer} D.~C.,  1990, \mn@doi [\apj] {10.1086/169195}, \href
  {http://adsabs.harvard.edu/abs/1990ApJ...361..300F} {361, 300}

\bibitem[\protect\citeauthoryear{{Genel} et~al.,}{{Genel}
  et~al.}{2014}]{genel2014}
{Genel} S.,  et~al., 2014, \mn@doi [\mnras] {10.1093/mnras/stu1654}, \href
  {http://adsabs.harvard.edu/abs/2014MNRAS.445..175G} {445, 175}

\bibitem[\protect\citeauthoryear{{Graham} et~al.,}{{Graham}
  et~al.}{2015}]{graham2015}
{Graham} M.~J.,  et~al., 2015, \mn@doi [\mnras] {10.1093/mnras/stv1726}, \href
  {http://adsabs.harvard.edu/abs/2015MNRAS.453.1562G} {453, 1562}

\bibitem[\protect\citeauthoryear{{Hellings} \& {Downs}}{{Hellings} \&
  {Downs}}{1983}]{hellings1983}
{Hellings} R.~W.,  {Downs} G.~S.,  1983, \mn@doi [\apjl] {10.1086/183954},
  \href {http://adsabs.harvard.edu/abs/1983ApJ...265L..39H} {265, L39}

\bibitem[\protect\citeauthoryear{{Hobbs} et~al.,}{{Hobbs}
  et~al.}{2010}]{hobbs2010}
{Hobbs} G.,  et~al., 2010, \mn@doi [Classical and Quantum Gravity]
  {10.1088/0264-9381/27/8/084013}, \href
  {http://adsabs.harvard.edu/abs/2010CQGra..27h4013H} {27, 084013}

\bibitem[\protect\citeauthoryear{Hunter}{Hunter}{2007}]{matplotlib2007}
Hunter J.~D.,  2007, Computing In Science \& Engineering, 9, 90

\bibitem[\protect\citeauthoryear{Jaffe \& Backer}{Jaffe \&
  Backer}{2003}]{jaffe2003}
Jaffe A.~H.,  Backer D.~C.,  2003, \mn@doi [The Astrophysical Journal]
  {10.1086/345443}, 583, 616

\bibitem[\protect\citeauthoryear{{Jenet}, {Hobbs}, {Lee}  \&
  {Manchester}}{{Jenet} et~al.}{2005}]{jenet2005}
{Jenet} F.~A.,  {Hobbs} G.~B.,  {Lee} K.~J.,   {Manchester} R.~N.,  2005,
  \mn@doi [\apjl] {10.1086/431220}, \href
  {http://adsabs.harvard.edu/abs/2005ApJ...625L.123J} {625, L123}

\bibitem[\protect\citeauthoryear{Jones, Oliphant, Peterson  et~al.}{Jones
  et~al.}{01  }]{scipy}
Jones E.,  Oliphant T.,  Peterson P.,   et~al., 2001--, {SciPy}: Open source
  scientific tools for {Python}, \url {http://www.scipy.org/}

\bibitem[\protect\citeauthoryear{{Kelley}, {Blecha}  \& {Hernquist}}{{Kelley}
  et~al.}{2017a}]{paper1}
{Kelley} L.~Z.,  {Blecha} L.,   {Hernquist} L.,  2017a, \mn@doi [\mnras]
  {10.1093/mnras/stw2452}, \href
  {http://adsabs.harvard.edu/abs/2017MNRAS.464.3131K} {464, 3131}

\bibitem[\protect\citeauthoryear{{Kelley}, {Blecha}, {Hernquist}, {Sesana}  \&
  {Taylor}}{{Kelley} et~al.}{2017b}]{paper2}
{Kelley} L.~Z.,  {Blecha} L.,  {Hernquist} L.,  {Sesana} A.,   {Taylor} S.~R.,
  2017b, \mn@doi [\mnras] {10.1093/mnras/stx1638}, \href
  {http://adsabs.harvard.edu/abs/2017MNRAS.471.4508K} {471, 4508}

\bibitem[\protect\citeauthoryear{{Kramer} \& {Champion}}{{Kramer} \&
  {Champion}}{2013}]{kramer2013}
{Kramer} M.,  {Champion} D.~J.,  2013, \mn@doi [Classical and Quantum Gravity]
  {10.1088/0264-9381/30/22/224009}, \href
  {http://adsabs.harvard.edu/abs/2013CQGra..30v4009K} {30, 224009}

\bibitem[\protect\citeauthoryear{{Lee}, {Wex}, {Kramer}, {Stappers}, {Bassa},
  {Janssen}, {Karuppusamy}  \& {Smits}}{{Lee} et~al.}{2011}]{lee201103}
{Lee} K.~J.,  {Wex} N.,  {Kramer} M.,  {Stappers} B.~W.,  {Bassa} C.~G.,
  {Janssen} G.~H.,  {Karuppusamy} R.,   {Smits} R.,  2011, \mn@doi [\mnras]
  {10.1111/j.1365-2966.2011.18622.x}, \href
  {http://adsabs.harvard.edu/abs/2011MNRAS.414.3251L} {414, 3251}

\bibitem[\protect\citeauthoryear{{Lentati} et~al.,}{{Lentati}
  et~al.}{2015}]{lentati2015}
{Lentati} L.,  et~al., 2015, \mn@doi [\mnras] {10.1093/mnras/stv1538}, \href
  {http://adsabs.harvard.edu/abs/2015MNRAS.453.2576L} {453, 2576}

\bibitem[\protect\citeauthoryear{{Lentati} et~al.,}{{Lentati}
  et~al.}{2016}]{lentati2016}
{Lentati} L.,  et~al., 2016, \mn@doi [\mnras] {10.1093/mnras/stw395}, \href
  {http://adsabs.harvard.edu/abs/2016MNRAS.458.2161L} {458, 2161}

\bibitem[\protect\citeauthoryear{{Magorrian} \& {Tremaine}}{{Magorrian} \&
  {Tremaine}}{1999}]{magorrian1999}
{Magorrian} J.,  {Tremaine} S.,  1999, \mn@doi [\mnras]
  {10.1046/j.1365-8711.1999.02853.x}, \href
  {http://adsabs.harvard.edu/abs/1999MNRAS.309..447M} {309, 447}

\bibitem[\protect\citeauthoryear{{Manchester} et~al.,}{{Manchester}
  et~al.}{2013}]{manchester2013a}
{Manchester} R.~N.,  et~al., 2013, \mn@doi [\pasa] {10.1017/pasa.2012.017},
  \href {http://adsabs.harvard.edu/abs/2013PASA...30...17M} {30, e017}

\bibitem[\protect\citeauthoryear{{McLaughlin}}{{McLaughlin}}{2013}]{mclaughlin2013}
{McLaughlin} M.~A.,  2013, \mn@doi [Classical and Quantum Gravity]
  {10.1088/0264-9381/30/22/224008}, \href
  {http://adsabs.harvard.edu/abs/2013CQGra..30v4008M} {30, 224008}

\bibitem[\protect\citeauthoryear{{Middleton}, {Chen}, {Del Pozzo}, {Sesana}  \&
  {Vecchio}}{{Middleton} et~al.}{2017}]{middleton2017}
{Middleton} H.,  {Chen} S.,  {Del Pozzo} W.,  {Sesana} A.,   {Vecchio} A.,
  2017, preprint, \href {http://adsabs.harvard.edu/abs/2017arXiv170700623M} {}
  (\mn@eprint {arXiv} {1707.00623})

\bibitem[\protect\citeauthoryear{{Nelson} et~al.,}{{Nelson}
  et~al.}{2015}]{nelson2015}
{Nelson} D.,  et~al., 2015, \mn@doi [Astronomy and Computing]
  {10.1016/j.ascom.2015.09.003}, \href
  {http://adsabs.harvard.edu/abs/2015A%26C....13...12N} {13, 12}

\bibitem[\protect\citeauthoryear{{Petiteau}, {Babak}, {Sesana}  \& {de
  Ara{\'u}jo}}{{Petiteau} et~al.}{2013}]{Petiteau2013}
{Petiteau} A.,  {Babak} S.,  {Sesana} A.,   {de Ara{\'u}jo} M.,  2013, \mn@doi
  [\prd] {10.1103/PhysRevD.87.064036}, \href
  {http://adsabs.harvard.edu/abs/2013PhRvD..87f4036P} {87, 064036}

\bibitem[\protect\citeauthoryear{{Phinney}}{{Phinney}}{2001}]{phinney2001}
{Phinney} E.~S.,  2001, ArXiv Astrophysics e-prints, \href
  {http://adsabs.harvard.edu/abs/2001astro.ph..8028P} {}

\bibitem[\protect\citeauthoryear{P�rez \& Granger}{P�rez \&
  Granger}{2007}]{ipython}
P�rez F.,  Granger B.,  2007, \mn@doi [Computing in Science Engineering]
  {10.1109/MCSE.2007.53}, 9, 21

\bibitem[\protect\citeauthoryear{{Rajagopal} \& {Romani}}{{Rajagopal} \&
  {Romani}}{1995}]{rajagopal1995}
{Rajagopal} M.,  {Romani} R.~W.,  1995, \mn@doi [\apj] {10.1086/175813}, \href
  {http://adsabs.harvard.edu/abs/1995ApJ...446..543R} {446, 543}

\bibitem[\protect\citeauthoryear{{Ravi}, {Wyithe}, {Hobbs}, {Shannon},
  {Manchester}, {Yardley}  \& {Keith}}{{Ravi} et~al.}{2012}]{ravi2012}
{Ravi} V.,  {Wyithe} J.~S.~B.,  {Hobbs} G.,  {Shannon} R.~M.,  {Manchester}
  R.~N.,  {Yardley} D.~R.~B.,   {Keith} M.~J.,  2012, \mn@doi [\apj]
  {10.1088/0004-637X/761/2/84}, \href
  {http://adsabs.harvard.edu/abs/2012ApJ...761...84R} {761, 84}

\bibitem[\protect\citeauthoryear{{Ravi}, {Wyithe}, {Shannon}, {Hobbs}  \&
  {Manchester}}{{Ravi} et~al.}{2014}]{ravi2014}
{Ravi} V.,  {Wyithe} J.~S.~B.,  {Shannon} R.~M.,  {Hobbs} G.,   {Manchester}
  R.~N.,  2014, \mn@doi [\mnras] {10.1093/mnras/stu779}, \href
  {http://adsabs.harvard.edu/abs/2014MNRAS.442...56R} {442, 56}

\bibitem[\protect\citeauthoryear{{Reardon} et~al.,}{{Reardon}
  et~al.}{2016}]{reardon2016}
{Reardon} D.~J.,  et~al., 2016, \mn@doi [\mnras] {10.1093/mnras/stv2395}, \href
  {http://adsabs.harvard.edu/abs/2016MNRAS.455.1751R} {455, 1751}

\bibitem[\protect\citeauthoryear{{Roebber}, {Holder}, {Holz}  \&
  {Warren}}{{Roebber} et~al.}{2016}]{roebber2016}
{Roebber} E.,  {Holder} G.,  {Holz} D.~E.,   {Warren} M.,  2016, \mn@doi [\apj]
  {10.3847/0004-637X/819/2/163}, \href
  {http://adsabs.harvard.edu/abs/2016ApJ...819..163R} {819, 163}

\bibitem[\protect\citeauthoryear{{Rosado}, {Sesana}  \& {Gair}}{{Rosado}
  et~al.}{2015}]{rsg15}
{Rosado} P.~A.,  {Sesana} A.,   {Gair} J.,  2015, \mn@doi [\mnras]
  {10.1093/mnras/stv1098}, \href
  {http://adsabs.harvard.edu/abs/2015MNRAS.451.2417R} {451, 2417}

\bibitem[\protect\citeauthoryear{{Sazhin}}{{Sazhin}}{1978}]{sazhin1978}
{Sazhin} M.~V.,  1978, Soviet Astronomy, \href
  {http://adsabs.harvard.edu/abs/1978SvA....22...36S} {22, 36}

\bibitem[\protect\citeauthoryear{{Schutz} \& {Ma}}{{Schutz} \&
  {Ma}}{2016}]{Schutz201510}
{Schutz} K.,  {Ma} C.-P.,  2016, \mn@doi [\mnras] {10.1093/mnras/stw768}, \href
  {http://adsabs.harvard.edu/abs/2016MNRAS.459.1737S} {459, 1737}

\bibitem[\protect\citeauthoryear{{Sesana}, {Haardt}, {Madau}  \&
  {Volonteri}}{{Sesana} et~al.}{2004}]{sesana2004}
{Sesana} A.,  {Haardt} F.,  {Madau} P.,   {Volonteri} M.,  2004, \mn@doi [\apj]
  {10.1086/422185}, \href {http://adsabs.harvard.edu/abs/2004ApJ...611..623S}
  {611, 623}

\bibitem[\protect\citeauthoryear{{Sesana}, {Vecchio}  \& {Colacino}}{{Sesana}
  et~al.}{2008}]{sesana2008}
{Sesana} A.,  {Vecchio} A.,   {Colacino} C.~N.,  2008, \mn@doi [\mnras]
  {10.1111/j.1365-2966.2008.13682.x}, \href
  {http://adsabs.harvard.edu/abs/2008MNRAS.390..192S} {390, 192}

\bibitem[\protect\citeauthoryear{{Sesana}, {Vecchio}  \& {Volonteri}}{{Sesana}
  et~al.}{2009}]{sesana2009}
{Sesana} A.,  {Vecchio} A.,   {Volonteri} M.,  2009, \mn@doi [\mnras]
  {10.1111/j.1365-2966.2009.14499.x}, \href
  {http://adsabs.harvard.edu/abs/2009MNRAS.394.2255S} {394, 2255}

\bibitem[\protect\citeauthoryear{{Sesana}, {Haiman}, {Kocsis}  \&
  {Kelley}}{{Sesana} et~al.}{2017}]{sesana201703}
{Sesana} A.,  {Haiman} Z.,  {Kocsis} B.,   {Kelley} L.~Z.,  2017, preprint,
  \href {http://adsabs.harvard.edu/abs/2017arXiv170310611S} {} (\mn@eprint
  {arXiv} {1703.10611})

\bibitem[\protect\citeauthoryear{{Shannon} et~al.,}{{Shannon}
  et~al.}{2015}]{shannon2015}
{Shannon} R.~M.,  et~al., 2015, \mn@doi [Science] {10.1126/science.aab1910},
  \href {http://adsabs.harvard.edu/abs/2015Sci...349.1522S} {349, 1522}

\bibitem[\protect\citeauthoryear{{Sijacki}, {Vogelsberger}, {Genel},
  {Springel}, {Torrey}, {Snyder}, {Nelson}  \& {Hernquist}}{{Sijacki}
  et~al.}{2015}]{sijacki2015}
{Sijacki} D.,  {Vogelsberger} M.,  {Genel} S.,  {Springel} V.,  {Torrey} P.,
  {Snyder} G.~F.,  {Nelson} D.,   {Hernquist} L.,  2015, \mn@doi [\mnras]
  {10.1093/mnras/stv1340}, \href
  {http://adsabs.harvard.edu/abs/2015MNRAS.452..575S} {452, 575}

\bibitem[\protect\citeauthoryear{{Taylor}, {Ellis}  \& {Gair}}{{Taylor}
  et~al.}{2014}]{taylor201406}
{Taylor} S.,  {Ellis} J.,   {Gair} J.,  2014, \mn@doi [\prd]
  {10.1103/PhysRevD.90.104028}, \href
  {http://adsabs.harvard.edu/abs/2014PhRvD..90j4028T} {90, 104028}

\bibitem[\protect\citeauthoryear{{Taylor}, {Huerta}, {Gair}  \&
  {McWilliams}}{{Taylor} et~al.}{2016a}]{taylor201505}
{Taylor} S.~R.,  {Huerta} E.~A.,  {Gair} J.~R.,   {McWilliams} S.~T.,  2016a,
  \mn@doi [\apj] {10.3847/0004-637X/817/1/70}, \href
  {http://adsabs.harvard.edu/abs/2016ApJ...817...70T} {817, 70}

\bibitem[\protect\citeauthoryear{{Taylor}, {Vallisneri}, {Ellis}, {Mingarelli},
  {Lazio}  \& {van Haasteren}}{{Taylor} et~al.}{2016b}]{taylor2016}
{Taylor} S.~R.,  {Vallisneri} M.,  {Ellis} J.~A.,  {Mingarelli} C.~M.~F.,
  {Lazio} T.~J.~W.,   {van Haasteren} R.,  2016b, \mn@doi [\apjl]
  {10.3847/2041-8205/819/1/L6}, \href
  {http://adsabs.harvard.edu/abs/2016ApJ...819L...6T} {819, L6}

\bibitem[\protect\citeauthoryear{{The NANOGrav Collaboration} et~al.,}{{The
  NANOGrav Collaboration} et~al.}{2015}]{arzoumanian2015a}
{The NANOGrav Collaboration} et~al., 2015, \mn@doi [\apj]
  {10.1088/0004-637X/813/1/65}, \href
  {http://adsabs.harvard.edu/abs/2015ApJ...813...65T} {813, 65}

\bibitem[\protect\citeauthoryear{{Torrey}, {Vogelsberger}, {Genel}, {Sijacki},
  {Springel}  \& {Hernquist}}{{Torrey} et~al.}{2014}]{torrey2014}
{Torrey} P.,  {Vogelsberger} M.,  {Genel} S.,  {Sijacki} D.,  {Springel} V.,
  {Hernquist} L.,  2014, \mn@doi [\mnras] {10.1093/mnras/stt2295}, \href
  {http://adsabs.harvard.edu/abs/2014MNRAS.438.1985T} {438, 1985}

\bibitem[\protect\citeauthoryear{{Verbiest} et~al.,}{{Verbiest}
  et~al.}{2016}]{Verbiest201602}
{Verbiest} J.~P.~W.,  et~al., 2016, \mn@doi [\mnras] {10.1093/mnras/stw347},
  \href {http://adsabs.harvard.edu/abs/2016MNRAS.458.1267V} {458, 1267}

\bibitem[\protect\citeauthoryear{{Vogelsberger}, {Genel}, {Sijacki}, {Torrey},
  {Springel}  \& {Hernquist}}{{Vogelsberger} et~al.}{2013}]{vogelsberger2013}
{Vogelsberger} M.,  {Genel} S.,  {Sijacki} D.,  {Torrey} P.,  {Springel} V.,
  {Hernquist} L.,  2013, \mn@doi [\mnras] {10.1093/mnras/stt1789}, \href
  {http://adsabs.harvard.edu/abs/2013MNRAS.436.3031V} {436, 3031}

\bibitem[\protect\citeauthoryear{{Vogelsberger} et~al.,}{{Vogelsberger}
  et~al.}{2014}]{vogelsberger2014b}
{Vogelsberger} M.,  et~al., 2014, \mn@doi [\mnras] {10.1093/mnras/stu1536},
  \href {http://adsabs.harvard.edu/abs/2014MNRAS.444.1518V} {444, 1518}

\bibitem[\protect\citeauthoryear{{Wyithe} \& {Loeb}}{{Wyithe} \&
  {Loeb}}{2003}]{wyithe2003}
{Wyithe} J.~S.~B.,  {Loeb} A.,  2003, \mn@doi [\apj] {10.1086/375187}, \href
  {http://adsabs.harvard.edu/abs/2003ApJ...590..691W} {590, 691}

\bibitem[\protect\citeauthoryear{{Yardley} et~al.,}{{Yardley}
  et~al.}{2011}]{yardley201102}
{Yardley} D.~R.~B.,  et~al., 2011, \mn@doi [\mnras]
  {10.1111/j.1365-2966.2011.18517.x}, \href
  {http://adsabs.harvard.edu/abs/2011MNRAS.414.1777Y} {414, 1777}

\bibitem[\protect\citeauthoryear{{Zhu} et~al.,}{{Zhu} et~al.}{2014}]{zhu2014}
{Zhu} X.-J.,  et~al., 2014, \mn@doi [\mnras] {10.1093/mnras/stu1717}, \href
  {http://adsabs.harvard.edu/abs/2014MNRAS.444.3709Z} {444, 3709}

\bibitem[\protect\citeauthoryear{{Zhu} et~al.,}{{Zhu} et~al.}{2015}]{zhu201502}
{Zhu} X.-J.,  et~al., 2015, \mn@doi [\mnras] {10.1093/mnras/stv381}, \href
  {http://adsabs.harvard.edu/abs/2015MNRAS.449.1650Z} {449, 1650}

\bibitem[\protect\citeauthoryear{{Zhu}, {Wen}, {Xiong}, {Xu}, {Wang},
  {Mohanty}, {Hobbs}  \& {Manchester}}{{Zhu} et~al.}{2016}]{zhu201606}
{Zhu} X.-J.,  {Wen} L.,  {Xiong} J.,  {Xu} Y.,  {Wang} Y.,  {Mohanty} S.~D.,
  {Hobbs} G.,   {Manchester} R.~N.,  2016, \mn@doi [\mnras]
  {10.1093/mnras/stw1446}, \href
  {http://adsabs.harvard.edu/abs/2016MNRAS.461.1317Z} {461, 1317}

\bibitem[\protect\citeauthoryear{van~der Walt, Colbert  \& Varoquaux}{van~der
  Walt et~al.}{2011}]{numpy2011}
van~der Walt S.,  Colbert S.~C.,   Varoquaux G.,  2011, CoRR, \href
  {https://arxiv.org/abs/1102.1523} {1102.1523}

\makeatother
\end{thebibliography}
